\documentclass[10pt]{article}

\usepackage{algorithm}
\usepackage[noend]{algpseudocode}
\usepackage{amsmath, amssymb, amsthm}
\usepackage{url}
\usepackage{fullpage}
\usepackage{prettyref}
\usepackage{pstricks, pst-node, graphicx}
\usepackage{xcolor}
\definecolor{colcite}{rgb}{0,0.75,0}
\definecolor{collink}{rgb}{0,0,1}
\definecolor{colurl}{rgb}{1,0,0}
\usepackage[colorlinks=true,citecolor=colcite,linkcolor=collink,urlcolor=colurl]{hyperref}

\newcommand{\Diam}{\ensuremath{\mathcal{D}}}

\newcommand{\Gr}{\ensuremath{G}}

\newcommand{\nequiv}{\not\equiv}

\newcommand{\Z}{\ensuremath{\mathbb{Z}}}

\newcommand{\bs}{\backslash}
\newcommand{\comment}[1]{}
\newcommand{\ignore}[1]{}
\newcommand{\Oh}[1]{O(#1)}

\newcommand{\rank}{\mathop{\mathrm{rank}}}
\newcommand{\dta}[1]{\mathop{\tt d}[#1]}
\newcommand{\w}{\mathop{\tt w}}
\newcommand{\algob}{\textsc{Rand-$b$-Bit-Circ}}
\newcommand{\xor}{\oplus}

\newcommand{\nspace}{}
\newcommand{\gcac}{\mathsf{Ca}(G)}

\newtheorem{theorem}{Theorem}[section]
\newtheorem{thm}[theorem]{Theorem}
\newtheorem{clm}[theorem]{Claim}

\newtheorem{prop}[theorem]{Proposition}
\newtheorem{lmma}[theorem]{Lemma}
\newtheorem{fact}[theorem]{Fact}
\newtheorem{cor}[theorem]{Corollary}
\theoremstyle{definition}
\newtheorem{defn}[theorem]{Definition}
\newrefformat{thm}{Theorem \ref{#1}}
\newrefformat{lmma}{Lemma \ref{#1}}
\newrefformat{ex}{Example \ref{#1}}
\newrefformat{clm}{Claim \ref{#1}}
\newrefformat{fact}{Fact \ref{#1}}
\newrefformat{cor}{Corollary \ref{#1}}
\newrefformat{conj}{Conjecture \ref{#1}}
\newrefformat{defn}{Definition \ref{#1}}
\newrefformat{prop}{Proposition \ref{#1}}
\newrefformat{app}{Appendix \ref{#1}}
\newrefformat{alg}{Algorithm \ref{#1}}
\newrefformat{line}{Line \ref{#1}}
\newrefformat{eq}{Equation \eqref{#1}}

\title{Fast Computation of Small Cuts via Cycle Space Sampling\footnote{A preliminary version of this work appeared as ``Fast Distributed Computation of Cuts via Random Circulations" in \emph{Proc.~35th ICALP} \cite{Pritchard08}.}}
\author{David Pritchard and Ramakrishna Thurimella\footnote{Email addresses: {\tt david.pritchard@epfl.ch}, {\tt ramki@cs.du.edu}}}

\begin{document}

\maketitle

\begin{abstract}
We describe a new sampling-based method to determine cuts
in an undirected graph. For a graph $(V, E)$, its cycle space is the family of all subsets of $E$
that have even degree at each vertex. We prove that with high probability, sampling the cycle space
identifies the cuts of a graph. This leads to simple new linear-time sequential algorithms
for finding all cut edges and \emph{cut pairs} (a set of 2 edges
that form a cut) of a graph.

In the model of \emph{distributed computing} in a graph $G=(V, E)$
with $O(\log |V|)$-bit messages, our approach yields faster
algorithms for several problems.
The diameter of $G$ is denoted by
$\Diam$, and the maximum degree by $\Delta$.
We obtain simple $O(\Diam)$-time distributed algorithms to find all
cut edges, 2-edge-connected components, and cut pairs, matching or
improving upon previous time bounds. Under natural
conditions these new algorithms are \emph{universally optimal} --- i.e.\ a $\Omega(\Diam)$-time lower bound holds on every graph. We obtain a $O(\Diam+\Delta/\log|V|)$-time
distributed algorithm for finding cut vertices; this is faster than
the best previous algorithm when $\Delta, \Diam =
O(\sqrt{|V|})$. A simple extension of our work
yields the first distributed algorithm with \emph{sub-linear} time
for 3-edge-connected components.  The basic distributed algorithms are Monte Carlo,
but they can be made Las Vegas without increasing the asymptotic
complexity.

In the model of \emph{parallel computing} on the EREW PRAM our approach yields a simple algorithm with optimal time complexity $O(\log V)$ for finding cut pairs and 3-edge-connected components.
\end{abstract}

\section{Introduction}\label{sec:intro}
Let $\Gr = (V, E)$ be a connected undirected graph. A part of $\Gr$
is said to be a \emph{cut} if, after deleting it from $\Gr$, the
remaining graph is disconnected. We use the following terminology:
\begin{itemize}
\item A \emph{cut vertex} is a vertex $v$ such that $\{v\}$ is a
cut.
\item A \emph{cut edge} is an edge $e$ such that $\{e\}$ is a
cut (i.e., a bridge).
\item A \emph{cut pair} is a cut consisting of two edges $e, f$ such that
neither $e$ nor $f$ is a cut edge.
\end{itemize}
For brevity we call all of these objects \emph{small cuts}. In a
network (e.g., for communication or transportation), the small cuts
are relevant because they represent the critical points where local
failures can cause global disruption. Our primary motivation is to
efficiently find all small cuts of an undirected graph. We study
this problem in the sequential, distributed, and parallel models of
computation.

The fundamentally new idea in this paper is to identify cuts by sampling the cycle space.
For a graph $(V, E)$ we say that $\phi \subseteq E$ is a \emph{binary circulation} if every vertex has even degree in $(V, \phi)$; the \emph{cycle space} of
graph $(V, E)$ is the set of all its binary circulations.
For $S \subseteq V$, let $\delta(S)$ denote the edges with exactly one
end in $S$. An \emph{induced edge cut} is a set of the form
$\delta(S)$ for some $S$; cut edges and cut pairs
are induced edge cuts\footnote{Our convention is that $\delta(\varnothing)=\delta(V)=\varnothing$ is an induced edge cut --- so we don't in general assume $\delta(S)$ is a cut.}. The family of all induced edge cuts is called the \emph{cut space} of a graph.
The cycle space and cut space are orthogonally complementary vector subspaces of $\Z_2^E$ (see \prettyref{sec:rc}), which implies that the intersection of any
binary circulation and induced edge cut is of even size.
At a high level, our
algorithms depend on a probabilistic converse (\prettyref{prop:nonsep-indep}): if $F\subset E$ is \emph{not} an induced edge cut, the number of edges of $F$ intersecting a uniformly random binary circulation is even with probability exactly 1/2. This specific observation seems new, although it is a simple consequence of standard results on the cut and cycle spaces. Using this observation we give efficient algorithms to sample a uniformly random binary circulation in the sequential, parallel, and distributed models of computing.

{\bf The Distributed Model.} Our approach improves several known time bounds
in the \emph{distributed} computing model with \emph{congestion}.
The precise model, denoted $\mathcal{CONGEST}$~\cite[\S
2.3]{p2000}, works as follows.
The computation takes place in the graph $G = (V, E)$ where each
vertex is a computer and each edge is a bidirectional communication
link; i.e., we study the problem of having a network compute the
small cuts of its own topology. There is no globally shared memory,
only local memory at each vertex. Initially only local topology is
known: each vertex knows its ID value, which is unique, and its
neighbours' IDs. Time elapses in discrete \emph{rounds}. In each
round, every vertex performs local computations and may send one
message to each of its
neighbors, to be received at the start of the next round. 
The \emph{time complexity} of a distributed
algorithm is the number of rounds that elapse, and the \emph{message
complexity} is the total number of messages that are sent.

In the $\mathcal{CONGEST}$ model, every message must be at most
$O(\log V)$ bits long. The model does not bound the memory capacity
or computational power of the vertices, although our algorithms use
time and space polynomial in $|V|$ at each vertex. Let $\Diam$
denote the diameter of $(V, E)$, i.e. $\Diam := \max_{u, v \in
V}\mathrm{dist}_G(u, v)$. The message size bound, in addition to
making the algorithms more practical, affects what is possible in
the model, as the following example from \cite{L+06} shows. On the one hand, if messages are allowed to
be arbitrarily long, any graph property whatsoever can be trivially
computed in $\Diam$ time\footnote{In $\Diam$ rounds each vertex
broadcasts its local topology to all other vertices, then each
vertex deduces the global topology and solves the problem with a
local computation.}. On the other hand, Lotker et al.\ gave a family
of graphs with $\Diam=3$, such that in $\mathcal{CONGEST}$ on this
family, a $\Omega(\sqrt[4]{|V|}/\sqrt{\log |V|})$-time lower bound
holds to find the minimum spanning tree (MST).

A distributed time complexity faster than $\Theta(V)$ on some graphs is called \emph{sub-linear}.
Determining whether a task in this model can be accomplished in
sub-linear time, or better yet $O(\Diam)$ time, is a
fundamental problem. E.g.\ one breakthrough was a sub-linear MST algorithm \cite{sublinear-mst} which was later improved \cite{kp-mst} to time complexity $O(\Diam+\sqrt{|V|}\log^* |V|)$ --- here $\log^* x$ is the number of times which $\log$
must be iteratively applied to $x$ before obtaining a number less
than 1. Our breakthroughs in this regard are $O(\Diam)$ time algorithms
for cut pairs, cut edges, and 2-edge-connected components, and a sub-linear algorithm for 3-edge-connected components.

\subsection{Existing Results}
Our results apply to three common models of computation: sequential,
distributed, and parallel. Abusing notation for readability, we
sometimes abbreviate $|V|$ to $V$ and $|E|$ to $E$.

{\bf Sequential.} In the usual sequential (RAM) model of computing,
Tarjan was the first to obtain linear-time
($O(V+E)$-time) algorithms to find all cut vertices
\cite{tarjandfs}, cut edges \cite{tarjandfs}, and cut vertex-pairs
(cuts $C \subseteq V$ with $|C|=2$) \cite{tri-tarjan}. These
algorithms are based on depth-first search (DFS). Galil \& Italiano \cite{galil-ital} gave the first linear-time algorithm to compute all cut
pairs, by reducing to the cut vertex-pair problem.

{\bf Distributed.} Here we only mention results valid in
$\mathcal{CONGEST}$, ignoring results with $\Omega(n)$ message size
such as one of \cite{chang}. {\bf Cut Edges/Vertices.} Two
early distributed algorithms for cut edges and vertices, in \cite{ahujazhu} and \cite{hohberg}, use DFS. The
smallest time complexity of any known distributed DFS algorithm is
$\Theta(V)$; as such, the algorithms of Ahuja \& Zhu and Hohberg
have $\Omega(V)$ time complexity. Huang \cite{huang} gave a
non-DFS-based algorithm with $\Theta(V)$ time complexity.
The first sub-linear distributed algorithms for any type of small cuts appear in \cite{Thur97}; using an MST subroutine, Thurimella obtained time complexity $O(\Diam+\sqrt{V}\log^* V)$ for both cut edges and cut vertices.
{\bf Cut Pairs.} For cut pairs,
\cite{JM96} gave a distributed algorithm with
worst-case time and message complexity $\Theta(n^3)$, and
\cite{2006tsin} obtained a DFS-based algorithm with
improved time complexity $O(\Diam^2+V)$.

{\bf Distributed Optimality.} Distributed $\Theta(V)$-time
algorithms for cut edges are optimal (up to a constant factor) on
some graphs: e.g.\ it is straightforward to see, even guaranteed
that $G$ is either a $|V|$-cycle or a $|V|$-path, not all edges can
determine if they are cut edges in less than $|V|/2-2$ rounds. One
term for this property is \emph{existentially optimal}, due to
\cite{sublinear-mst}. However, as
Thurimella's algorithm \cite{Thur97} showed, there are some graphs
on which $\Theta(V)$ time is not asymptotically optimal. The
stronger term \emph{universally optimal}~\cite{sublinear-mst}
applies to algorithms which, on \emph{every} graph, have running
time within a constant factor of the minimum possible.

{\bf Parallel.} In the PRAM model,
an optimal $O(\log V)$-time and
$O(V+E)$-work Las Vegas algorithm for cut edges and vertices was obtained in \cite{tarjan-parallel} (provided that for spanning forests, recent work of
\cite{HZ01} is used). For cut pairs, it may be possible to combine a
a 3-vertex-connectivity algorithm of \cite{FRT93} with the reduction of \cite{galil-ital} (and spanning forest routines from \cite{HZ01}) to yield a time- and work-optimal EREW algorithm.
This is mentioned as a ``future application" by Halperin \& Zwick. However, this approach appears not to have been fully analyzed and is fairly complicated.

\subsection{Our Contributions}\label{sec:contributions}
Since our algorithms are randomized, we differentiate between two
types of algorithms: \emph{Monte Carlo} ones have deterministically
bounded running time but may be incorrect with probability at most $1/V$ and
\emph{Las Vegas} ones are always correct and have bounded
\emph{expected} running time\footnote{More generally, our algorithms
can obtain error probability $\leq 1/V^c$ for any constant $c$
without changing the asymptotic complexity.}. (Note, a Las Vegas
algorithm can always be converted to Monte Carlo, so Las Vegas is
generally better).

{\bf Sequential.} The random circulation approach yields \emph{new
linear-time algorithms to compute all cut edges and cut pairs} of
the Las Vegas type. As far as we are aware, our linear-time cut pair
algorithm is the first one that does not rely on either DFS (e.g.,
see references in \cite{2005tsin}) or open ear decomposition
(e.g., see references in \cite{FRT93}).

{\bf Distributed.} We remark that all existing distributed
algorithms mentioned for finding small cuts are deterministic. The
random circulation approach yields \emph{faster distributed
algorithms for small cuts} of the Las Vegas type. For cut edges and
pairs, we obtain $O(\Diam)$-time algorithms. Compared to the
previous best time of $O(\Diam+\sqrt{V}\log^* V)$ for cut edges, we
remove the dependence on $|V|$. Compared to the previous best   time
of $O(\Diam^2+V)$ for cut pairs, we obtain a quadratic speedup on
every graph. For cut vertices, we obtain a $O(\Diam+\Delta/\log
V)$-time algorithm where $\Delta$ is the maximum degree. Compared to
the previous best time of $O(\Diam+\sqrt{V}\log^* V)$ for cut
vertices, this is faster on graphs with $\Delta, \Diam =
O(\sqrt{V})$. We also obtain the first sub-linear distributed
algorithm for 3-edge-connected components, using a connected
components subroutine of \cite{Thur97}. In Table
\ref{tab1} we depict our main results and earlier work, showing both
time and message complexity.

{\bf Universal Optimality.} If we assume distributed algorithms must
act globally in a natural sense --- either by initiating at a single
vertex, or by reporting termination
--- then a $\Omega(\Diam)$-time lower bound holds for the problems
of finding cut edges or cut pairs, on any graph. Hence under natural
conditions, our $O(\Diam)$-time algorithms for cut edges and cut
pairs are universally optimal.

{\bf Parallel.} In the PRAM model, we obtain a Las Vegas algorithm for cut pairs and 3-edge-connected components
with time complexity $O(\log V+T(E))$, space complexity $O(E+S(E))$, and work complexity $O(E + W(E))$, where $T(n), S(n), W(n)$ are respectively the time, space, work complexity to sort $n$ numbers of length $O(\log n)$ bits. E.g.\ on the EREW PRAM, we can implement our algorithm in $O(\log V)$ time, $O(E)$ space and $O(E \log E)$ work using a sorting subroutine of \cite{KRS90}, or in $O(\log V)$ time, $O(E^{1+\epsilon})$ space and $O(E \sqrt{\log E})$ work using a subroutine of \cite{HS02}.

\begin{table}[ht]
\begin{center}
\begin{tabular}{ccccc}
\hline \hline
 &  \phantom{X} Cuts Found \phantom{X} & Time & Messages\\
\hline
 \cite{ahujazhu} &  Vertices \& Edges  & $\Oh{V}$ & $\Oh{E}$  \\
 \cite{Thur97} &  Vertices \& Edges & $\Oh{\Diam +
\sqrt{V} \log^* V}$ & $\Oh{E \cdot (\Diam +
\sqrt{V} \log^* V)}$ \\
 \cite{2006tsin}  & Pairs & $\Oh{V+\Diam^2}$ & $\Oh{E+V\cdot\Diam}$  \\
\prettyref{thm:dce}\dag & Edges & $\Oh{\Diam}$ &
$\Oh{E}$  \\
\prettyref{thm:dcpair}\dag & Pairs &
$\Oh{\Diam}$ &
$\Oh{\min\{V^2, E \cdot \Diam\}}$  \\
\prettyref{thm:dcv}\dag & Vertices & $\Oh{\Diam
+ \Delta / \log V}$ & $\Oh{E (1 +\Delta / \log V )}$
\\ \hline \hline
\end{tabular} \caption{Comparison of our three main distributed results (denoted by \dag) to the best previously known algorithms. \nspace}\label{tab1}
\end{center}
\end{table}

\subsection{Other Related Work}
Randomized algorithms appear in other literature related to the cut and cycle spaces. For example, \cite{BL03} computes the genus of an
embedded graph $G$ while ``observing" part of it. They use
random perturbation and balancing steps to compute a \emph{near-circulation} on $G$
and the dual graph of $G$. Their computational model is quite different from the one here, e.g.\
they allow a face to modify the values of all its incident edges in a single time step.

A slow bridge-finding algorithm based on random walks is given in \cite{PV06}, which inspires this paper. Random sampling is a fruitful technique to quickly compute so-called \emph{minimum cycle bases} of the cycle space, e.g.~see the survey~\cite{KL+09}.

\subsection{Organization of the Paper}
In \prettyref{sec:rc} we define random
circulations and show how to construct them efficiently. In
\prettyref{sec:app} we show how random circulations yield algorithms
for small cuts and give sequential implementations. In \prettyref{sec:impl} we
precisely define the assumptions in our distributed model and give the Monte Carlo algorithms; we introduce a technique called
\emph{fundamental cycle-cast} which may be of independent interest. In \prettyref{sec:cc} we discuss 2- and 3-edge-connected components. In \prettyref{sec:lasvegas} we give the Las Vegas analogues of our distributed algorithms.  We give $\Omega(\Diam)$ distributed time lower bounds under precise assumptions in \prettyref{sec:lowerbounds}. We give the parallel cut pair algorithm in \prettyref{sec:parallel}.

\section{Preliminaries on Circulations} \label{sec:rc}
The cut space and cycle space over $\Z$ in directed graphs have been studied for quite some time \cite{BM76}. For our purposes it is convenient to work modulo 2; then, informally, we can deal with undirected graphs since $+1 \equiv -1 \pmod{2}$. For the sake of completeness, we prove the needed results. See also \cite{Diestel} which proves material equivalent to Propositions \ref{prop:vs}, \ref{prop:orth}, and \ref{prop:coolio}.

For notational convenience we identify any subset $S$ of $E$ with its \emph{characteristic vector} $\chi^S \in \Z_2^E$ defined by $\chi^S_e = 1$ for $e \in S$ and $\chi^S_e = 0$ for $e \not\in S$. We use $\xor$ to stand for vector addition modulo 2, so in accordance with our notational convention, for $S, T \subset E$ the expression $S \xor T$ denotes the symmetric difference of $S$ and $T$.

As mentioned earlier, $\phi \subseteq E$ is a \emph{binary circulation} if in $(V, \phi)$ every vertex has even degree; the \emph{cycle space} of graph $(V, E)$ is the set of all its binary circulations; $\delta(S)$ denotes the edges of $G$ with exactly one
end in $S$; an \emph{induced edge cut} is a set of the form $\delta(S)$ for some $S$; and the family of all induced edge cuts is called the \emph{cut space} of a graph. For $v \in V$ we use $\delta(v)$ as short for $\delta(\{v\})$.

\begin{prop}\label{prop:vs}
The cut space and cycle space are vector subspaces of $\Z_2^E$.
\end{prop}
\begin{proof}
Note it suffices to show each space contains $\varnothing$ and is closed under $\xor$. For the cut space, this holds since $\delta(\varnothing)=\varnothing$ and $\delta(S \xor T) = \delta(S) \xor \delta(T)$. For the cycle space, clearly $(V, \varnothing)$ has even degree at each vertex; and if $(V, \phi_1)$ and $(V, \phi_2)$ have even degree at each vertex, then the degree of vertex $v$ in $(V, \phi_1 \xor \phi_2)$ is $\deg_{\phi_1}(v) + \deg_{\phi_2}(v) - 2\deg_{\phi_1 \cap \phi_2}(v) \equiv 0 + 0 - 0 \pmod{2}$, so $\phi_1 \xor \phi_2$ is a binary circulation.
\end{proof}

\begin{prop}\label{prop:orth}
The cut space and cycle space are orthogonal.
\end{prop}
\begin{proof}
We need precisely to show that for any binary circulation $\phi$ and any $S \subset V$ that the dot product $\phi \cdot \delta(S) \equiv 0 \pmod{2}$, or equivalently that $|\phi \cap \delta(S)|$ is even. Now $\sum_{s \in S} \deg_\phi(s) = \sum_{s \in S} |\phi \cap \delta(s)|$ and the former quantity is even since $\phi$ is a circulation. The latter sum counts every edge of $\phi \cap \delta(S)$ once, every edge of $\phi$ with both ends in $S$ twice, and every other edge zero times. Since this sum is even, $|\phi \cap \delta(S)|$ is even.
\end{proof}

In the next proposition, we assume $G$ is connected, and hence has a spanning tree $T$. We need to define the \emph{fundamental cuts} and \emph{fundamental cycles} of $T$. For each edge $e$ of $E \bs E(T)$, we define the \emph{fundamental cycle} $C_e$ to be the unique cycle in $T \cup \{e\}$. Note cycles are binary circulations. For each edge $e$ of $T$, we define $S_e$ to be one of the two connected components of $T \bs e$, and define the \emph{fundamental cut of $e$} to be $\delta(S_e)$ (note $\delta(S_e)$ does not depend on which connected component we chose).

\begin{prop}\label{prop:coolio}
(a) The cut space and cycle space are orthogonal complements. (b) The cycle space has dimension $|E|-|V|+1$ and the cut space has dimension $|V|-1$. (c) For any spanning tree $T$ of $G$, its fundamental cycles form a basis of the cycle space, and its fundamental cuts form a basis of the cut space.\end{prop}
\begin{proof}
We will show that the $|E|-|V|+1$ fundamental cycles are linearly independent in the cycle space and the $|V|-1$ fundamental cuts are linearly independent in the cut space. Basic linear algebra shows the sum of the dimensions of two orthogonal subspaces of $\Z_2^E$ is at most $|E|$, with equality only if they are orthogonal complements, thus by \prettyref{prop:orth}, \prettyref{prop:coolio}(a) and (b) follow, and so does (c). We use the following claim.
\begin{clm}\label{clm:tech}
Let $H \subset E$ and consider a family of vectors $\{x^e\}_{e \in H}$ over $\Z_2^E$. If $x^e_e = 1$ for all $e \in H$, and $x^e_f = 0$ for all distinct $e, f \in H$, then $\{x^e\}_{e \in H}$ is linearly independent.
\end{clm}
\begin{proof}
Suppose for the sake of contradiction that $\bigoplus_{e \in H} a_e x^e$ is the zero vector, where $a_e \in \{0,1\}$ for each $e$ and not all $a_e$ are zero. Pick $f$ such that $a_f=1$, then $\sum_{e \in H} a_e x^e_f = 1,$ a contradiction.
\end{proof}
Note that $e \in C_e$ but for any other edge $f$ of $E \bs E(T)$, $f \not\in C_e$, so by \prettyref{clm:tech} with $H=E \bs E(T)$ and $x^e = C_e$, these vectors are linearly independent. Note that $e \in \delta(S_e)$ but for any other edge $f$ of $T$, $f \not\in \delta(S_e)$, so by \prettyref{clm:tech} with $H=E(T)$ and $x^e = \delta(S_e)$, these vectors are linearly independent. This completes the proof of \prettyref{prop:coolio}.
\end{proof}

\subsection{Random Circulations}\label{sec:randcirc}
Next we show why uniform sampling of the cycle space is useful for identifying cuts.
\begin{prop}\label{prop:nonsep-indep}
Let $F \subset E$ be a set that is not an induced edge cut. If $\phi$ is a uniformly random binary circulation, then $\Pr[|F \cap \phi| \textrm{ is even}] = 1/2.$
\end{prop}
\begin{proof}
Since $F$ is not in the cut space, by \prettyref{prop:coolio}(a) it is not orthogonal to the cycle space, i.e.\ there is a binary circulation $\phi_F$ with $|F \cap \phi_F|$ odd. Now we pair up each binary circulation $\psi$ on $G$ with the binary circulation $\psi' := \psi \xor \phi_F$. This yields a pairing of all binary circulations on $G$ since for all $\psi$, $\psi' \neq \psi$ and $\psi'' = \psi$. Modulo 2, $|F \cap \psi'| \equiv |F \cap \psi| + |F \cap \phi_F| \equiv |F \cap \psi|+1$, so in each pair, exactly one of the two binary circulations has even intersection with $F$. Thus, exactly half of all binary circulations have even intersection with $F$, which proves the result.
\end{proof}

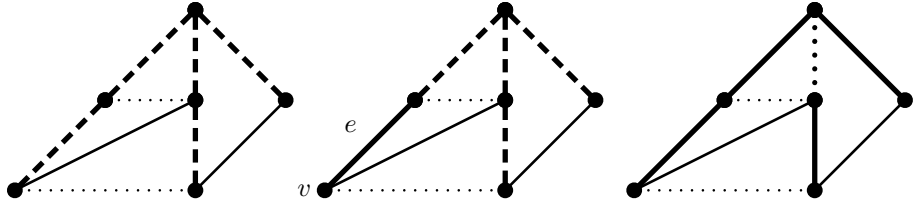
\begin{figure}[t]
\begin{center} \leavevmode
\begin{pspicture}(0,0)(4, 2.4)
  \psset{unit=1.2}
  \pnode(0,0){a}
  \pnode(2,0){b}
  \pnode(1,1){c}
  \pnode(2,1){d}
  \pnode(3,1){e}
  \pnode(2,2){f}
  \psset{linewidth=2pt,linestyle=dashed}
  \ncline{*-*}{f}{c}
  \ncline{*-*}{f}{d}
  \ncline{*-*}{f}{e}
  \ncline{*-*}{c}{a}
  \ncline{*-*}{d}{b}
  \psset{linewidth=1pt}
  \psset{linestyle=solid}
  \ncline{*-*}{b}{e}
  \ncline{*-*}{a}{d}
  \psset{linestyle=dotted}
  \ncline{*-*}{c}{d}
  \ncline{*-*}{a}{b}
\end{pspicture}
\begin{pspicture}(0,0)(4, 2.4)
  \psset{unit=1.2}
  \pnode(0,0){a}\uput[180](0,0){$v$}
  \pnode(2,0){b}
  \pnode(1,1){c}
  \pnode(2,1){d}
  \pnode(3,1){e}
  \pnode(2,2){f}
  \psset{linewidth=2pt,linestyle=dashed}
  \ncline{*-*}{f}{c}
  \ncline{*-*}{f}{d}
  \ncline{*-*}{f}{e}
  \ncline{*-*}{d}{b}
  \psset{linewidth=2pt,linestyle=solid}
  \ncline{*-*}{c}{a}\Bput{$e$}
  \psset{linewidth=1pt}
  \ncline{*-*}{a}{d}
  \ncline{*-*}{b}{e}
  \psset{linewidth=1pt,linestyle=dotted}
  \ncline{*-*}{a}{b}
  \ncline{*-*}{c}{d}
\end{pspicture}
\begin{pspicture}(0,0)(4, 2.4)
  \psset{unit=1.2}
  \pnode(0,0){a}
  \pnode(2,0){b}
  \pnode(1,1){c}
  \pnode(2,1){d}
  \pnode(3,1){e}
  \pnode(2,2){f}
  \psset{linewidth=2pt,linestyle=solid}
  \ncline{*-*}{f}{c}
  \ncline{*-*}{f}{e}
  \ncline{*-*}{d}{b}
  \ncline{*-*}{c}{a}
  \psset{linewidth=2pt,linestyle=dotted}
  \ncline{*-*}{f}{d}
  \psset{linewidth=1pt,linestyle=solid}
  \ncline{*-*}{a}{d}
  \ncline{*-*}{b}{e}
  \psset{linewidth=1pt,linestyle=dotted}
  \ncline{*-*}{a}{b}
  \ncline{*-*}{c}{d}
\end{pspicture}
\end{center}
\caption{Completing a binary circulation. The spanning tree $T$ is given by thick edges. Solid edges are in the circulation, dotted edges will not be in the circulation, and dashed edges are undecided. Left: the initial value of $\phi$ (which equals $\psi$). Middle: we ensure a leaf vertex $v$ has even degree. Right: repeating the previous step yields the completed circulation $\phi$.} \label{fig:example0}
\end{figure}

Next we give a method for constructing binary circulations (it is an undirected version of \cite[Ex.\ 12.1.1]{BM76}). Given a spanning tree $T$ and subset $\psi$ of $E \bs E(T)$, we say that $\phi$ is a \emph{completion} of $\psi$ if $\phi$ is a binary circulation and $\phi \cap (E \bs E(T)) = \psi$.
\begin{prop} \label{prop:cotree-basis}
For any $\psi \subseteq E \bs E(T)$, $\psi$ has a unique completion $\phi$.
\end{prop}
\begin{proof}
First, we give a succinct proof sketch. By \prettyref{prop:coolio}(c) the cycle space can be expressed as $\{\bigoplus_{e \in E \bs E(T)} a_e C_e \mid a \in \Z_2^{E \bs E(T)}\}$. For which $a$ does this yield a completion of $\psi$? From the observations in the proof of \prettyref{prop:coolio}, for $f \in E \bs E(T)$, the coordinate of $\bigoplus_{e \in E \bs E(T)} a_e C_e$ at index $f$ is just $a_f$, hence the unique completion of $\psi$ is the one in which $a$ is the indicator vector of $\psi$, i.e.\ the unique completion is $\phi = \bigoplus_{e \in \psi} C_e$. Explicitly, for $f\in T$, we have $f \in \phi$ iff $f$ appears in an odd number of the fundamental cycles $\{C_e \mid e \in \psi\}$. This completes the proof, but we now give a second, algorithmic proof, which is needed later.

For a leaf node $v$ incident to $e \in E(T)$, since the degree
of $(V, \phi)$ at $v$ must be even, notice that we must have $e \in \phi$
if $|\psi \cap \delta(v)|$ is odd, and $e \not\in \phi$
if $|\psi \cap \delta(v)|$ is even. Iterating this argument on $T \bs v$ yields \prettyref{alg:complete};
we will show it
constructs the unique completion of $\psi$.

\begin{algorithm}[ht]
\caption{Given $G, T$ and $\psi \subset E \bs E(T)$, construct binary circulation $\phi$ such that $\phi \bs E(T) = \psi$.}\label{alg:complete}
\begin{algorithmic}[1]
\State Initialize $\phi := \psi,
S := T$ \Comment{$S$ is the subtree of $T$ where $\phi$ is not yet
defined}
\State {\bf while} $S$ has any edges,
\State \quad Let $v$ be any leaf of $S$ and $e$ be the unique incident edge of $v$ in $S$
\State \quad {\bf if} $|\delta(v) \cap \phi|$ is odd {\bf then} $\phi := \phi \cup \{e\}$
\label{line:conserv} \Comment{Satisfy degree constraint at $v$}
\label{line:0flow}
\State \quad Delete $v$ from $S$
\State Output $\phi$
\end{algorithmic}
\end{algorithm}

See Figure~\ref{fig:example0} for an illustration of \prettyref{alg:complete}. Now we prove \prettyref{prop:cotree-basis} using \prettyref{alg:complete}. It is clear that every vertex of $(V, \phi)$ has even degree except possibly the last vertex left in $S$. However,
by the handshake lemma, no graph can have exactly one vertex of odd degree, so $\phi$ is indeed a binary circulation.
To show uniqueness, suppose for the sake of contradiction that $\psi$ has two distinct completions $\phi, \phi'$. Then $\phi \xor \phi' \subset E(T)$, and as such the nonempty forest $\phi \xor \phi'$ has at least one vertex of degree 1. This contradicts the fact that $\phi \xor \phi'$ is a binary circulation.
\end{proof}

We now give the method
for constructing uniformly random binary circulations, illustrated in \prettyref{alg:rc}: pick a uniformly random subset of $E \bs E(T)$ and then compute its completion.

\begin{algorithm}[ht]
\caption{Given $G$ and spanning tree $T$, output a uniformly random binary circulation.}\label{alg:rc}
\begin{algorithmic}[1]
\State {\bf for} each $e$ in $E \bs E(T)$, put $e$ in $\psi$ with independent probability 1/2 \label{line:rcast}
\State Return the completion of $\psi$, using \prettyref{alg:complete}
\end{algorithmic}
\end{algorithm}

\begin{thm}\label{thm:urand}
\prettyref{alg:rc} outputs a uniformly random binary circulation.
\end{thm}
\begin{proof}By \prettyref{prop:coolio}(b) the cycle space contains exactly $2^{|E|-|V|+1}$ elements. \prettyref{alg:rc} makes one of $2^{|E|-|V|+1}$ choices of $\psi$ each with probability $2^{-|E|+|V|-1}$, and each distinct choice of $\psi$ leads to a distinct binary circulation.\end{proof}

\begin{figure}[t]
\begin{center} \leavevmode
\begin{pspicture}(0,0)(4, 2.4)
  \psset{unit=1.2}
  \pnode(0,0){a}
  \pnode(2,0){b}
  \pnode(1,1){c}
  \pnode(2,1){d}
  \pnode(3,1){e}
  \pnode(2,2){f}
  \psset{linewidth=2pt}
  \ncline{*-*}{f}{c}
  \ncline{*-*}{f}{d}
  \ncline{*-*}{f}{e}
  \ncline{*-*}{c}{a}
  \ncline{*-*}{d}{b}
  \psset{linewidth=1pt}
  \ncline{*-*}{c}{d}\mput*{010}
  \ncline{*-*}{a}{d}\mput*{100}
  \ncline{*-*}{a}{b}\mput*{011}
  \ncline{*-*}{b}{e}\mput*{111}
\end{pspicture}
\begin{pspicture}(0,0)(4, 2.4)
  \psset{unit=1.2}
  \pnode(0,0){a}\uput[180](0,0){$v$}
  \pnode(2,0){b}
  \pnode(1,1){c}
  \pnode(2,1){d}
  \pnode(3,1){e}
  \pnode(2,2){f}
  \psset{linewidth=2pt}
  \ncline{*-*}{f}{c}
  \ncline{*-*}{f}{d}
  \ncline{*-*}{f}{e}
  \ncline{*-*}{c}{a}\mput*{111}\Bput{$e$}
  \ncline{*-*}{d}{b}
  \psset{linewidth=0.8pt}
  \ncline{*-*}{c}{d}\mput*{010}
  \ncline{*-*}{a}{d}\mput*{100}
  \ncline{*-*}{a}{b}\mput*{011}
  \ncline{*-*}{b}{e}\mput*{111}
\end{pspicture}
\begin{pspicture}(0,0)(4, 2.4)
  \psset{unit=1.2}
  \pnode(0,0){a}
  \pnode(2,0){b}
  \pnode(1,1){c}
  \pnode(2,1){d}
  \pnode(3,1){e}
  \pnode(2,2){f}
  \psset{linewidth=2pt}
  \ncline{*-*}{f}{c}\mput*{101}
  \ncline{*-*}{f}{d}\mput*{010}
  \ncline{*-*}{f}{e}\mput*{111}
  \ncline{*-*}{c}{a}\mput*{111}
  \ncline{*-*}{d}{b}\mput*{100}
  \psset{linewidth=0.8pt}
  \ncline{*-*}{c}{d}\mput*{010}
  \ncline{*-*}{a}{d}\mput*{100}
  \ncline{*-*}{a}{b}\mput*{011}
  \ncline{*-*}{b}{e}\mput*{111}
\end{pspicture}
\end{center}
\caption{Constructing a random 3-bit circulation; thick edges are tree edges and thin edges are non-tree edges. Left: we assign random $\phi$ values to the non-tree edges. Middle: we set $\phi(e) := \bigoplus_{f \in \delta(v) \bs e} \phi(f)$ for a leaf vertex $v$. Right: repeating the previous step yields the completed circulation $\phi$.} \label{fig:example}
\end{figure}
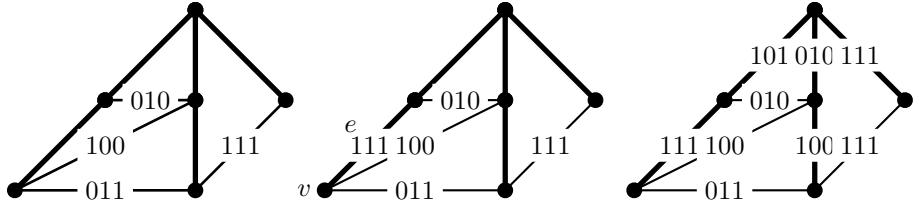

To increase the probability of identifying a particular cut beyond 1/2, our algorithms will sample multiple independent random  circulations. For this reason it is convenient to introduce notation that incorporates multiple circulations into a single object.
Let $\Z_2^b$ denote the set of $b$-bit binary strings. For $\phi: E
\to \Z_2^b$, let $\phi_i(e)$ denote the $i$th bit of $\phi(e)$.
\begin{defn}
$\phi: E \to \Z_2^b$ is a \emph{$b$-bit circulation} if for each $1 \leq i \leq b$, $\{e \mid \phi_i(e)=1\}$ is a binary  circulation.
\end{defn}
Hence, to say that $\phi$ is a uniformly random $b$-bit circulation is the same as saying that $\{\phi_i\}_{i=1}^b$ are mutually independent, uniformly random binary circulations. For brevity, we use the phrase \emph{random $b$-bit circulation} to stand for ``uniformly random $b$-bit circulation" in the rest of the paper.
Let {\bf 0} denote the all-zero vector and $\oplus$ denote addition of vectors modulo 2. Using \prettyref{prop:orth} and \prettyref{prop:nonsep-indep} we obtain the following corollary.
\begin{cor}\label{cor:nonsep-indep}
Let $\phi$ be a random $b$-bit circulation and $F \subseteq E$. Then
$$\Pr\left[\bigoplus_{e \in F} \phi(e) = {\bf 0}\right] = \begin{cases}1, &\textrm{ if $F$ is an induced edge cut;} \\ 2^{-b}, &\textrm{ otherwise.}\end{cases}$$
\end{cor}
To generate a random $b$-bit circulation, it suffices to modify Algorithms \ref{alg:complete} and \ref{alg:rc} slightly so as to operate independently on each of $b$ positions at once: on  \prettyref{line:rcast} of \prettyref{alg:rc} we set $\phi(e)$ to a uniformly independent $b$-bit string, and on \prettyref{line:conserv}
of \prettyref{alg:complete} we set $\phi(e) := \bigoplus_{f \in \delta(v) \bs e} \phi(f)$. We denote the resulting algorithm by \algob\ and illustrate it in Figure \ref{fig:example}.
Under the standard assumption that the machine word size is $\Theta(\log V)$, the running time of \algob\ in the sequential model of computing is $O(E\lceil \frac{b}{\log V} \rceil)$.

\section{Basic Algorithms}\label{sec:app}
In this section we show how to use random circulations to
probabilistically determine the cut edges, cut pairs, and cut
vertices of a graph. These are the Monte Carlo versions of the
algorithms.

\subsection{Finding All Cut Edges}\label{sec:cutedgealg}
We provide
pseudocode in \prettyref{alg:edge} and then prove its correctness. It is based on the easy fact that $e$ is a cut edge if and only if $\{e\}$ is an induced edge cut, which we state without proof.

\begin{algorithm}[ht]
\caption{Given a connected graph $\Gr,$ compute the cut edges of $\Gr.$}\label{alg:edge}
\begin{algorithmic}[1]
\State Let $b = \lceil \log_2 VE \rceil$ and let $\phi$ be a random $b$-bit circulation
on $\Gr$. \State Output all edges $e$ for which $\phi(e)={\bf 0}$ \label{line:edgecheck}
\end{algorithmic}
\end{algorithm}
\nspace

\begin{thm}\prettyref{alg:edge} correctly determines the cut edges with probability at least $1-1/V$ and can be implemented in $O(E)$
sequential time.\label{thm:sedge}
\end{thm}
\begin{proof}
Using the fact above, \prettyref{cor:nonsep-indep}, and a union bound, the probability of error
is at most $E/2^b \leq 1/V$. The subroutine \algob\, as well as \prettyref{line:edgecheck} of \prettyref{alg:edge}, each take $O(E)$ sequential time.
\end{proof}

\subsection{Finding All Cut Pairs and Cut Classes}\label{sec:cutpairs}
\prettyref{prop:foo}, whose easy proof we omit, leads to our approach for finding cut
pairs.
\begin{prop}[Cut pairs are induced]\label{prop:foo}
Let $e$ and $f$ be edges that are not cut edges. Then $\{e, f\}$ is a cut pair if and only if $\{e, f\}$ is an induced edge cut.
\end{prop}
\ignore{\begin{proof}
Since $e$ is not a cut edge $G \bs \{e\}$ is connected, and so $G
\bs \{e, f\}$ must have exactly two connected components. Let $U$ be
the vertex set of one of them, and so the other is $V \bs U$.

Note, $f$ must be incident on both $U$ and $V \bs U$; indeed,
otherwise $G \bs \{e\}$ would not be connected. A similar claim
holds for $e$; i.e., both $e$ and $f$ lie in $\delta(U).$ No other
edges can lie in $\delta(U)$ since this would contradict the fact
that $U$ is a connected component of $G \bs \{e, f\}$. Hence
$\delta(U) = \{e, f\}$. \end{proof}}
With \prettyref{cor:nonsep-indep} we immediately obtain the following.
\begin{cor} \label{cor:seppair}
Let $e, f$ be two distinct edges that are not cut edges. Then $\Pr[\phi(e) = \phi(f)] =
1$ if $\{e, f\}$ is a cut pair, and $2^{-b}$ otherwise.
\end{cor}
This yields a cute probabilistic proof of the following basic fact.
\begin{cor}[Transitivity of cut pairs] \label{cor:cutpairclass}
If $\{e, f\}$ and $\{f, g\}$ are cut pairs, then so is $\{e, g\}$.
\end{cor}
\begin{proof}
Note that $e, f, g$ are not cut edges. Let $\phi$ be a random
1-bit circulation on $\Gr$. By \prettyref{cor:seppair},
$\phi(e)=\phi(f)$ and $\phi(f)=\phi(g)$.  So $\phi(e)=\phi(g)$ with
probability 1. By \prettyref{cor:seppair}, $\{e, g\}$ must be a cut
pair. \end{proof}

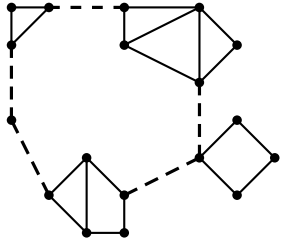
\begin{figure}[t]
\begin{center} \leavevmode
\begin{pspicture}(-1.5,-0.6)(3,2.1)
\psset{unit=0.5cm}
\psline[linewidth=1.2pt,linestyle=dashed](-2,3)(-2,1)
\psline[linewidth=1.2pt,linestyle=dashed](-2,1)(-1,-1)
\psline(-1,-1)(0,-2)
\psline(0,-2)(1,-2)
\psline(1,-2)(1,-1)
\psline(1,-1)(0,0)
\psline(-1,-1)(0,0)
\psline(0,0)(0,-2)
\psline[linewidth=1.2pt,linestyle=dashed](1,-1)(3,0)
\psline[linewidth=1.2pt,linestyle=dashed](3,0)(3,2)
\psline(3,0)(4,1)
\psline(4,1)(5,0)
\psline(5,0)(4,-1)
\psline(4,-1)(3,0)
\psline(3,2)(4,3)
\psline(4,3)(3,4)
\psline(3,4)(1,4)
\psline(1,4)(1,3)
\psline(1,3)(3,2)
\psline(3,4)(1,3)
\psline(3,2)(3,4)
\psline[linewidth=1.2pt,linestyle=dashed,dash=4pt 4pt](1,4)(-1,4)
\psline(-1,4)(-2,4)
\psline(-2,4)(-2,3)
\psline(-2,3)(-1,4)
\psdots(-2,3)
\psdots(-1,-1)
\psdots(0,-2)
\psdots(1,-2)
\psdots(1,-1)
\psdots(0,0)
\psdots(3,0)
\psdots(3,2)
\psdots(4,1)
\psdots(5,0)
\psdots(4,-1)
\psdots(4,3)
\psdots(3,4)
\psdots(1,4)
\psdots(1,3)
\psdots(-1,4)
\psdots(-2,4)
\psdots(-2,1)
\end{pspicture}
\end{center}
\caption{A graph is shown with one cut class highlighted using dashed edges. Deleting any two dashed edges disconnects the graph.} \label{fig:cc}
\end{figure}

\begin{defn}A \emph{cut class} is an inclusion-maximal subset $K$ of
$E$ such that $|K|>1$ and every pair $\{e, f\} \subseteq K$ is a cut
pair.\end{defn} We illustrate a cut class in Figure~\ref{fig:cc}. Note the cut class has a natural cyclic order.

\prettyref{cor:cutpairclass} implies that any
two distinct cut classes are disjoint. Hence, even though there may
be many cut pairs, we can describe them all compactly by listing all cut classes
of the graph.
We now give our simple linear-time algorithm to find all cut
classes\ignore{The idea is to compute a random $b$-bit circulation for
large enough $b$ that $\phi(e)={\bf 0}$ only for cut edges, and so
that $\phi$ labels the cut classes of other edges.}, with pseudocode given in \prettyref{alg:pair}.

\begin{algorithm}[ht]
\caption{Given a connected graph $\Gr,$ compute the cut classes of $\Gr.$}\label{alg:pair}
\begin{algorithmic}[1]
\State Let $b = \lceil \log_2 (VE^2) \rceil$ and let $\phi$ be a
random $b$-bit circulation on $\Gr$
\label{line:sort} \State {\bf for} each $x \in \Z_2^b \bs \{\bf 0\}$ such that
$|\{e \in E \mid \phi(e) = x\}| \geq 2,$ output the cut class
$\{e \in E \mid \phi(e) = x\}$\label{line:val-loop}
\end{algorithmic}
\end{algorithm}
\nspace
\begin{thm}\prettyref{alg:pair} correctly determines the cut pairs with probability at least $1-1/V$ and can be implemented in $O(E)$
sequential time.\label{thm:cpair}\end{thm}
\begin{proof}
There are $|E|$ edges and the analysis in \prettyref{sec:cutedgealg} shows that
$\Pr[\phi(e) = {\bf 0}] \leq 1/2^b$ for each non-cut edge $e$. There
are at most $\tbinom{E}{2}$ pairs $\{e, f\}$ of non-cut edges that
are not cut pairs and \prettyref{cor:seppair} shows that
$\Pr[\phi(e) = \phi(f)] \leq 1/2^b$ for each such pair. Hence, by a
union bound,
the total probability of error is at most $E/2^b+\tbinom{E}{2}/2^b
\leq 1/V$.

The subroutine \algob\ has time complexity $O(E)$. It remains to implement \prettyref{line:val-loop} of
\prettyref{alg:pair} in $O(E)$ time. To do this, we sort all edges $e$ according to the key
$\phi(e)$ using a three-pass \emph{radix sort}.
I.e., we consider each value in $\Z_2^b$ as a three-digit number in
base $2^{b/3} = O(E)$ --- see \cite[\S
9.3]{CLR90} --- then the sort takes $O(E)$ time. \end{proof}

\subsection{Finding All Cut Vertices}\label{sec:cutvertexalg}
The following characterization of cut vertices underlies our approach.
\begin{prop}\label{prop:cvchar}
The cut $\delta(v)$ properly contains a nonempty induced edge cut if and only if $v$ is a cut vertex.
\end{prop}
\begin{proof}
First, suppose $v$ is a cut vertex. Let $V_1$ be the vertex set of one of the connected components of
$\Gr \bs \{v\}.$ Then $\delta(v)$ properly contains the nonempty induced edge cut $\delta(V_1)$.

Second, suppose $v$ is not a cut vertex, so there is a spanning tree $T'$ of $G \bs \{v\}$. Suppose $S \subset V$ has $\delta(S) \subseteq \delta(v)$. Without loss of generality (by complementing $S$ if necessary) we assume $v \in S$. Since no edges of $T'$ are in $\delta(S)$, $S$ either contains all of $V \bs \{v\}$ or none of $V \bs \{v\}$. Thus either $S = V$ in which case $\delta(S)$ is empty, or $S = \{v\}$, in which case $\delta(S)$ is not a proper subset of $\delta(v)$.
\end{proof}
Using \prettyref{prop:cvchar}, the essential
idea in our approach to find cut vertices is to detect for each vertex $v$ whether $\delta(v)$ properly contains any nonempty induced edge cuts. As usual we detect induced edge cuts via
\prettyref{cor:nonsep-indep}, this time rephrasing the detection
problem as one of finding linearly dependent rows of a
matrix. Hence we need the following fact, when $\Z_2$ is viewed as a
field.
\begin{fact} \label{fact:binsum} In a matrix over $\Z_2,$ a set $C$ of columns is
linearly dependent if and only if some nonempty subset of $C$ sums
to the zero column vector $(\bmod\,2)$.
\end{fact}

Our approach works as follows. Note --- it does not have a very efficient sequential implementation, but yields an efficient distributed algorithm.
We generate a random $b$-bit
circulation $\phi$ for some suitably large $b$; denote the $i$th bit
of $\phi(e)$ by $\phi_i(e)$. Let $d(v) := |\delta(v)|$, the
\emph{degree} of $v$. Let $\Delta$ denote the maximum degree. For
each vertex $v$, let $M^{[v]}$ be a matrix with $b$ rows indexed $1,
\dotsc, b$, and $d(v)$ columns indexed by $\delta(v)$; then fill the
entries of $M^{[v]}$ according to $M^{[v]}_{ie} = \phi_i(e)$. The
following two complementary claims validate our approach.
\begin{clm} \label{clm:sleep} If $v$ is a cut vertex then
$\rank(M^{[v]}) \leq d(v)-2$.
\end{clm}
\begin{proof}
Let $V_1$ be the vertex set of one of the connected components of
$\Gr \bs \{v\}.$ Note that $\delta(v)$ can be partitioned into two
induced edge cuts $\delta(V_1)$ and $\delta(\{v\} \cup V_1).$ By
\prettyref{cor:nonsep-indep} the set of columns of $M^{[v]}$ corresponding
to $\delta(V_1)$ adds to zero, and by \prettyref{fact:binsum} these
columns are linearly dependent. Similarly, the remaining columns, indexed by
$\delta(\{v\} \cup V_1)$, are linearly dependent. So $M^{[v]}$ has at
least 2 columns that are linearly dependent on the others, and the
result follows.\end{proof}

\begin{clm} \label{clm:apa} Let $v \in V$ and assume that $v$ is
not a cut vertex. Let $\varnothing \subsetneq D \subsetneq \delta(v).$
The probability that the columns of $M^{[v]}$ indexed by $D$ sum to
the zero vector $(\bmod\,2)$ is $2^{-b}.$
\end{clm}
\begin{proof}
By \prettyref{prop:cvchar}, $D$ is not an induced edge cut, and the result follows from \prettyref{cor:nonsep-indep}.
\end{proof}

Next we show that for $b = \lceil \Delta + 2 \log_2 V \rceil$, it
is very likely that $\rank(M^{[v]})<d(v)-1$ iff $v$ is a cut vertex.
Thus our approach, with pseudocode given in \prettyref{alg:vertex},
is correct with high probability.

\begin{algorithm}[ht]
\caption{Given a connected graph $\Gr,$ compute the cut vertices of $\Gr.$}\label{alg:vertex}
\begin{algorithmic}[1]
\State Let $b = \lceil \Delta + 2 \log_2 V \rceil$ and let $\phi$
be a random $b$-bit circulation on $G$ \State {\bf for} each vertex $v$
of $\Gr$, {\bf if} $\rank(M^{[v]}) < d(v)-1$ {\bf then} output $v$
\end{algorithmic}
\end{algorithm}
\nspace

\begin{thm}\prettyref{alg:vertex} correctly determines the cut vertices with probability at least
$1-1/V$. \label{thm:svert} \end{thm} \begin{proof}
\prettyref{clm:sleep} shows that all cut vertices are output.
Consider a vertex $v$ that is not a cut vertex and let $D$ be a
subset of $\delta(v)$ of size $d(v)-1$. By \prettyref{clm:apa},
\prettyref{fact:binsum}, and a union bound, the probability that the
columns of $M^{[v]}$ corresponding to $D$ are linearly dependent is
at most $2^{d(v)-1}2^{-b} \leq 1/V^2;$ so with probability at least
$1-V^{-2},$ we have $\rank(M^{[v]}) \geq |D| = d(v)-1$ and $v$ is
not output. By another union bound, the probability that any vertex
is misclassified by \prettyref{alg:vertex} is at most
$V/V^2=1/V.$ \end{proof}

\section{Distributed Implementation}\label{sec:impl}
Our algorithms make the following three assumptions: first, the
network is synchronous; second, there is a distinguished
\emph{leader} vertex at the start of computation; third, every node
begins with a unique $O(\log V)$-bit ID. These assumptions are
standard in the sense that they are made by the best previous
distributed algorithms \cite{ahujazhu,Thur97,2006tsin} for small
cuts. Nonetheless, these assumptions can be removed at a cost if
desired, e.g. using the synchronizer of \cite{synch} at a polylog($V$) factor increase in complexity,
Peleg's \cite{pelegopt} $O(\Diam)$-time leader election algorithm,
or by randomly assigning IDs in the range $\{1, \dotsc, V^3\}$
(resulting in additional failure probability at most
$\tbinom{V}{2}/V^3$ due to ID collisions).

Although only vertices can store data in the distributed model, we
maintain data for each edge $e$ (e.g., to represent a tree) by
having both endpoints of $e$ store the data. At the end of the
algorithm, we require that the correct result is known locally, so
each node stores a boolean variable indicating whether it is a cut
node, and similarly for edges. To indicate cut pairs, each edge must
know whether it is in any cut pair, and in addition we must give
every cut class a distinct label. Previous work also essentially
uses these representations.

When stating distributed algorithms, the assumptions of a leader,
synchrony, unique IDs, and $O(\log V)$-bit messages are implicit.
Our algorithms use a breadth-first search (BFS) tree with a root $r$
as the basis for communication. One reason that BFS trees are useful
is that they can be constructed quickly (e.g., see \cite[\S
5.1]{p2000}), as follows.
\begin{prop}\label{prop:bfs}
There is a distributed algorithm to construct a BFS tree in
$O(\Diam)$ time and $O(E)$ messages.
\end{prop}
\noindent For a tree $T$, the \emph{level} $l(v)$ of $v \in V$ is
the distance in $T$ between $v$ and $r$. The \emph{height} $h(T)$ of
tree $T$ is the maximum vertex level in $T$. Any BFS tree $T$ has
$h(T) \leq \Diam$ and this is important because several fundamental
algorithms based on passing information up or down the tree take
$O(h(T))$ time. The \emph{parent} of $u$ is denoted $p(u)$. The
\emph{level of tree edge $\{u, p(u)\}$} is the level of $u.$

\subsection{Random Circulations and Cut Edges}
 \label{sec:cutedgeimpl}
When we construct a random circulation, we require at termination
that each $v$ knows $\phi(e)$ for each $e \in \delta(v)$.
\begin{thm}\label{thm:drc}\label{thm:zdrc}
There is a distributed algorithm to sample a random
$b$-bit circulation in $O(\Diam)$ time and $O(E)$ messages, when $b = O(\log V)$.
\end{thm}
\begin{proof}
We implement \algob\
distributively. The size bound ensures that $b$-bit strings can be sent in a message.
We compute a BFS
tree $T$, using \prettyref{prop:bfs}. Then for each non-tree edge
$\{e\}$ in parallel, the endpoint with the higher ID
picks a random $b$-bit value for $\phi(e)$ and sends it to the other endpoint.
In the following $h(T)$ rounds, for $i=h(T)$ down to 1,
each level-$i$ vertex computes $\phi(\{v, p(v)\}) := \bigoplus_{f \in \delta(v) \bs \{v, p(v)\}} \phi(f)$ and sends
this value to $p(v)$.
The complexity is $O(\Diam+h(T))=O(\Diam)$ time and $O(E+E)$
messages. \end{proof}

\prettyref{thm:drc} yields our distributed cut edge algorithm.
\begin{thm}\label{thm:dce}
There is a distributed algorithm to compute all cut edges with
probability at least $1-1/V$ in $O(\Diam)$ time and using $O(E)$ messages.
\end{thm}
\begin{proof}
We implement \prettyref{alg:edge} distributively, obtaining the
required correctness probability by \prettyref{thm:sedge}. For $k =
VE$, we use \prettyref{thm:drc} to compute a random
$\lceil \log_2 VE \rceil$-bit circulation in the required complexity bounds. Then we identify
$e$ as a cut edge if $\phi(e)={\bf 0}$. \end{proof}

\subsection{Pipelining and Cut Vertices}
Our cut vertex algorithm requires a circulation on $\Theta(\Delta+\log V)$ bits, and in order to construct such a circulation quickly, we use a \emph{pipelining} technique. Let $\pi$ be a distributed algorithm in which for each edge $e$, the
total number of messages sent on $e$ by $\pi$ is bounded by some
universal constant $C_0$. The messages' content may be random but
the message-passing schedule must be deterministic. To
\emph{pipeline $s$ instances of $\pi$} means to execute $s$
instances $\{\pi_i\}_{i=1}^s$ of $\pi$, each one delayed by a unit
time step from the previous. When multiple instances need to
simultaneously send messages along the same edge we concatenate
them, increasing the message sizes by a factor of at most $C_0$.
Compared to $\pi$, pipelining adds $s-1$ to the time complexity and
increases the message complexity by a factor of $s.$

A straightforward implementation of \prettyref{alg:vertex} results
in our cut vertex algorithm, as follows.
\begin{thm}\label{thm:dcv}
There is a distributed algorithm to compute all cut vertices with
probability at least $1-1/V$ in $O(\Diam + \Delta / \log V)$ time and using
$O(E (1+ \Delta / \log V ))$ messages.
\end{thm}
\begin{proof}
We implement \prettyref{alg:vertex}
distributively, obtaining probability $1/V$ of failure by
\prettyref{thm:svert}. Let $b = \lceil
\Delta + 2 \log_2 V \rceil.$
\prettyref{thm:zdrc} gives an algorithm $\pi$
to construct a random $O(\log V)$-bit circulation; note $\pi$ sends
a constant number of messages along each edge. We pipeline $b/\log V$ instances of $\pi$ to construct a random $b$-bit
circulation. Then, each vertex $v$ locally computes the rank of
$M^{[v]}$ to determine if it is a cut vertex.

Since $\pi$ takes
$O(\Diam)$ rounds and sends $O(E)$ messages, and $b = O(\Delta +
\log V),$ the implementation takes $O(\Diam + \Delta / \log V)$ time
and $O(E (1+ \Delta / \log V ))$ messages.\end{proof}

\subsection{Fundamental Cycle-Cast (fc-cast)} \label{sec:fccast}\label{sec:fc}
We now define a new distributed technique, needed for our cut pair algorithm. A \emph{non-tree edge} is an
edge $e \in E \bs E(T)$. For a spanning tree $T$ and non-tree edge
$e,$ the unique cycle in $T \cup \{e\}$ is called \emph{the
fundamental cycle of $T$ and $e$}, and we denote it by $C_e$. We
call our new technique \emph{fundamental cycle-cast}, or
\emph{fc-cast} for short, and informally it allows simultaneous
processing on all fundamental cycles. Let each vertex $v$ store some
data $\dta{v}$ of length $O(\log V)$ bits. We assume that $\dta{v}$ includes
the ID, level, and parent ID of $v$, since this information can be appended to $\dta{v}$ while increasing its length by at most $O(\log V)$ bits.
At the end of the fc-cast, each
non-tree edge $e$ will know $\dta{u}$ for every vertex $u$ in the
fundamental cycle of $T$ and $e.$

\begin{thm}\label{thm:fc-cast} There is a distributed algorithm {\sc Fc-Cast} using $O(h(T))$
time and $O(\min\{E\cdot h(T), V^2\})$ messages that, for each
non-tree edge $e$, for each $v \in C_e$, sends $\dta{v}$ to both
endpoints of $e$.
\end{thm}

As a subroutine, we need a tree broadcast subroutine adapted from \cite[\S 3.2]{p2000}.
\begin{prop}\label{prop:tree-cast} There is a distributed algorithm {\sc Tree-Broadcast} using $O(h(T))$
time and $O(V \cdot h(T))$ messages that sends $\dta{v}$ to $u$ for
each $v \in V$ and each descendant $u$ of $v$. \end{prop}
\begin{proof}
Let $\pi$ be a generic distributed algorithm that sends one message
from $p(v)$ to $v$ at time $l(v);$ in particular, $\pi$ takes $O(V)$
messages, $O(h(T))$ time, and sends at most one message on each
edge. Define instances $\{\pi_i\}_{i=0}^{h(t)}$ of $\pi$ so that for
every vertex $v$ at level $i$, and for every descendant $u$ of $v$,
instance $\pi_i$ is responsible for propagating $\dta{v}$ to $u$.
Each instance $\pi_i$ sends empty messages for the first $i$ rounds,
and in round $t > i$, for each $v$ with $l(v)=i$, propagates
$\dta{v}$ down the level-$t$ tree edges descending from $v$. Since
there are $h(T)+1$ pipelined instances and $\pi$ takes $O(h(T))$
time and $O(V)$ messages, the complexity follows. \end{proof}

\begin{proof}[Proof of \prettyref{thm:fc-cast}]
An fc-cast has two steps. First, we execute {\sc Tree-Broadcast},
and as a result we may assume that each vertex has a \emph{list} of
the data of all its ancestors.

In the second step, for each non-tree edge $\{v, w\}$ in parallel,
$v$ sends its list to $w$ and vice-versa. Note that each non-tree
edge $e$ can determine its fundamental cycle with $T$ by comparing
its endpoints' lists. (More precisely, either endpoint of $e$ can
determine such.) Each list has at most $1+h(T)$ items, each of which
is $O(\log V)$ bits long and can be sent in a single message, so
both steps in the fc-cast take $O(h(T))$ time.

The message
complexity of the second step as just described is $O(E\cdot h(T))$,
but now we give a refinement that achieves $O(\min\{E\cdot h(T),
V^2\})$ message complexity.
The essential idea is for all $u, v \in V$, we want to avoid sending
$\dta{u}$ to $v$ more than once. Implement the second step of the
fc-cast so that each vertex $v$ sends one $\dta{\cdot}$ value per
round, and in the order $\dta{v}$ first, then $\dta{p(v)},$ etc.,
with the data of the root last. When a vertex $u$ receives $\dta{x}$
for the second time for some $x$, $u$ asks the sender to stop
sending its list. Likewise, if $u$ receives $\dta{x}$ from multiple
neighbors at the same time, $u$ asks all but one to stop sending
their lists. Along each edge, at most one redundant message and one
stop request can be sent in each direction. There can only be
$V^2$ non-redundant messages; hence the total number of messages
sent in this step is $O(V^2+E)$. Considering the tree-broadcast as
well, the total message complexity is $O(V \cdot h(T) + \min\{E\cdot
h(T), V^2 + E\}) = O(\min\{E\cdot h(T), V^2\})$ as claimed.
\end{proof}

We can implement fc-cast in $O(h(T))$ time with message complexity even smaller than $\min\{E\cdot h(T), V^2\}$ using a nearest common ancestor labeling scheme of \cite{AGKR02}. We only sketch the idea since the precise improved complexity is somewhat awkward to state (seemingly cannot be expressed in terms of parameters $V, E, \Delta, h(T)$) and does not seem universally optimal. If $uw$ is an edge not in $T$, call $w$ a \emph{non-tree neighbour} of $u$ and vice-versa. The general idea behind the optimized implementation is that, while the implementation in \prettyref{thm:fc-cast} sends $\dta{v}$ to each descendant of $v$ and each non-tree neighbour of a descendant of $v$, we can actually send $\dta{v}$ to a smaller subset of these nodes while meeting the definition of a fundemental cycle-cast.

In more detail, the scheme of \cite{AGKR02} gives each vertex an $O(\log V)$-bit label such that given \emph{just the labels} of any two nodes, we can also compute the label of their \emph{nearest common ancestor} (with a deterministic algorithm independent of $T$). Alstrup et al.\ do not work in any specific distributed model, but their scheme is built out of  standard primitives like the number of descendants of a given node, and as such can be implemented in the model we consider in $O(h(T))$ time and $O(E)$ messages. The first step of our new implementation is to compute these labels. Then, in unit time and $2|E|$ messages, we have each node inform each of its neighbours of its label.

At a high level, the labeling scheme allows the implementation to be optimized as follows. In the first step we send $\dta{v}$ down to its descendant $u$ only if there is some fundamental cycle containing both $u$ and $v$; in the second step each $v$ asks for $\dta{\cdot}$ values from its non-tree neighbours in such a way that $u$ receives each $\dta{\cdot}$ value at most once, and only asks for $\dta{v}$ from $w$ if $C_{uw}$ contains $v$. Implementing these steps requires that nodes have some knowledge about the relative position of their neighbours in the tree, which is accomplished using the labels. There are some slightly complicated details in implementing the first step, for which a pipelined \emph{convergecast} (see \prettyref{prop:con-cast}) suffices.

\subsection{Distributed Cut Pair Algorithm}\label{sec:cutpairimpl}
When computing the cut pairs,
it helps if we assume that
$\Gr$ has no cut edges, i.e.\ $G$ is 2-edge-connected.
To make this assumption without loss of generality, for our input graph $G$, we compute the set $E_C$
of cut edges using \prettyref{thm:lvedge} and then report the cut pairs
of the \emph{2-edge-connected components}, which are the connected components of $G \bs E_C$ (we elaborate in \prettyref{sec:cc}).
It is straightforward to show that the cut pairs of $G$ are the cut pairs of these components, that each component has no cut edge,
and that no component has diameter greater than $G$.

It is not obvious how to implement our sequential cut pair algorithm
(\prettyref{alg:pair}) distributively: although the cut classes are
properly labeled with high probability by $\phi$, in order for edge
$e$ to know whether it belongs to any cut pair, it needs to
determine if any other $f$ has $\phi(e)=\phi(f)$, and this cannot be
done using local information (i.e., in $O(1)$ rounds). We use
fc-cast to overcome this obstacle. The following claims are used to relate
fundamental cycles to cut classes. (The first is fairly intuitive given \prettyref{fig:cc}.)
\begin{lmma} \label{lmma:ccc} If a cycle $C$ and a cut class $K$ satisfy $K \cap C \neq \varnothing$ then $K
\subseteq C.$
\end{lmma}
\begin{proof}
Suppose that $e \in K \cap C$ but $f \in K \bs C.$ Then by
\prettyref{prop:foo}, $\{e, f\}$ is an induced edge cut. But then $|\{e, f\} \cap C|=1$,
contradicting \prettyref{prop:orth} (the orthogonality of the cut space and cycle space).
\end{proof}

\begin{clm}\label{clm:responsible}
Let $K$ be a cut class. Then $K \subset C_e$ for some $e \in E \bs
E(T)$.
\end{clm}
\begin{proof} First we claim $K$ contains at most one non-tree edge. Suppose otherwise, for the sake of contradiction, that $K$
contains two non-tree edges $\{e, f\}$. Then $\{e, f\}$ is a cut pair and so $\Gr \bs \{e, f\}$
is not connected. However, this contradicts the fact that $\Gr \bs \{e, f\}$ contains the spanning tree $T$.

The definition of a cut class implies $|K|>1$, so $K$ contains at least one
tree edge $e$. Since $e$ is not a cut edge, $\Gr \bs \{e\}$ is
connected, and hence there is a non-tree edge $f$ that connects the two
connected components of $T \bs \{e\}.$ The fundamental cycle $C_f$ of $f$
and $T$ thus contains $e,$ and by \prettyref{lmma:ccc}, all of $K.$
\end{proof}

To describe our cut pair algorithm we introduce a variant of a
standard technique, the \emph{convergecast} (e.g., see
\cite[\S 4.2]{p2000}). Informally, it allows each node to
independently query its descendants. In this paper we take the convention that $v$ is always a descendant of itself. Let $Desc(v)$ denote the set
of $v$'s descendants. For each $v \in V$, and
each $u \in Desc(v)$, let $\w[u, v]$ be a variable of length $\Theta(\log V)$ stored
at $u$.
\begin{prop}\label{prop:con-cast}
There is a distributed algorithm {\sc Converge-Cast} that uses $O(h(T))$
time and $O(V \cdot h(T))$ messages so that each $v \in V$
determines $\max \{\w[u, v] \mid u \in Desc(v)\}.$
\end{prop}
\begin{proof}
We assume some familiarity with the basic implementation of convergecast in order to gloss over some basic details; see
\cite[\S 4.2]{p2000}.
We use $\pi$ to represent a generic distributed algorithm that sends messages from leaves to the root in level-synchronized fashion. The ``standard" convergecast uses $\pi$ to compute $\max \{\w[u, r] \mid u \in V\}$ at $r$; in round $i$, for $i$ from $h(T)$ down to 1, every level-$i$ node passes up the largest value that it knows about to its parent.
A slight modification yields instances $\{\pi_i\}_{i=0}^{h(t)}$ of $\pi$ so
that for every vertex $v$ at level $i$, instance $\pi_i$ propagates $\max \{\w[u, v] \mid u \in Desc(v)\}$ to
$v$.
Since there are $h(T)+1$ pipelined instances and $\pi$ takes
$O(h(T))$ time and $O(V)$ messages, the complexity follows.
\end{proof}

\ignore{Finally, we explain the details of convergecast.
\begin{proof}[Proof of \prettyref{prop:con-cast}: {\sc Converge-Cast}]
Let $\pi$ be a generic distributed algorithm that sends one message
from $v$ to $p(v)$ at time $1+h(T)-l(v);$ in particular, $\pi$ takes
$O(V)$ messages, $O(h(T))$ time, and sends at most one message on
each edge. Define instances $\{\pi_i\}_{i=0}^{h(t)}$ of $\pi$ so
that for every vertex $v$ at level $i$, instance $\pi_i$ is
responsible for propagating $\bigvee_{u \in Desc(v)} \w[u, v]$ to
$v$.

We implement $\pi_i$ as follows. For $v' \in Desc(v)$, define
$x[v', v] := \bigvee_{u \in Desc(v')} \w[u, v]$. Each level-$h(T)$
vertex $v'$ can immediately compute $x[v', v]$ for all its ancestors
$v$. In $h(T)$ rounds, for $j$ from $h(T)$ down to 1, for each
vertex $v'$ at level $j$ and each ancestor $v$ of $v'$ at level $i$,
$v'$ computes $x[v', v]$ and sends $x[v', v]$ to $p(v')$. Observe
that
$$x[v', v] = \w[v', v] \vee \bigvee_{v'' \textrm{ a child of $v'$}}
x[v'', v].$$ Hence, $x[v', v]$ can be computed by $v'$ by taking the
OR of $\w[v', v]$ and all values sent up to $v'$ from its children
in the previous round.

Since there are $h(T)+1$ pipelined instances and $\pi$ takes
$O(h(T))$ time and $O(V)$ messages, the complexity follows.
\end{proof}}

\begin{thm}\label{thm:dcpair}
There is a distributed algorithm to compute all cut classes with
probability at least $1-1/V$ in $O(\Diam)$ time and using $O(\min\{E\cdot
\Diam, V^2\})$ messages.
\end{thm}
\begin{proof} As in \prettyref{alg:pair}, for
$b = \lceil \log_2 (VE^2) \rceil$ we compute a random $b$-bit
circulation $\phi$ on $\Gr$, using \prettyref{thm:zdrc}. Denote the
following assumption by \eqref{eq:ass2}.
\comment{\begin{equation}\textrm{For all edges $e$, $\phi(e)={\bf
0}$ if and only if $e$ is a cut edge}. \tag{\ensuremath{\star}}
\label{eq:ass1}\end{equation} \vspace{-0.7cm}}
\begin{equation} \textrm{For all edges $e, f$, $\phi(e)=\phi(f)$ if and only if $\{e, f\}$ is a cut pair}.
\tag{\ensuremath{\star}} \label{eq:ass2}\end{equation} \noindent  By
the analysis in the proof of \prettyref{thm:cpair}, we may assume that
\eqref{eq:ass2} holds without violating the required bound of $1/V$
on the probability of error.

It remains only for each edge to determine whether it is a member of
any cut pair, since then $\phi$ labels the cut classes. For each
vertex $v \neq r$ let $\dta{v} := \phi(\{v, p(v)\}).$ We run {\sc
Fc-Cast}, and as a result, the endpoints of each non-tree edge $e$
can compute the multiset $\Phi_e := \{\phi(f) \mid f \in C_e\}$. The
following claim, which follows immediately from \prettyref{clm:responsible}, lets each non-tree edge determine if it is a member
of any cut pair.
\begin{clm}\label{clm:phish1}
A non-tree edge $e$ is in a cut pair if and only if $\phi(e)$ occurs
multiple times in $\Phi_e$.
\end{clm}
To deal with tree edges, for each $v \in V$ and each $u \in
Desc(v)$, define
$$\w[u, v] := | \{ e \in \delta(u) \bs E(T) \mid \{v, p(v)\} \in C_e \textrm{ \&  $\phi(\{v, p(v)\})$ occurs $\ge 2$
times in }\Phi_e \} |.$$ \noindent and note that $\w[u, v]$ can be
determined by $u$ after the fc-cast. We run {\sc Converge-Cast}.

\begin{clm}\label{clm:phish2}
Tree edge $\{v, p(v)\}$ is in a cut pair if and only if $\exists u
\in Desc(v)$ such that $\w[u, v]>0$.
\end{clm}
\begin{proof}
If $\{v, p(v)\}$ lies in a cut pair then by \prettyref{clm:responsible} there is a fundamental cycle $C_e$ containing that cut pair. It is easy to see that one endpoint $u$ of $e$ is a descendant of $v$ and has $\w[u, v]>0$.
\end{proof}

By \prettyref{prop:con-cast}, after the convergecast, each tree edge
can use \prettyref{clm:phish2} to determine if it is a member of any
cut pair. Adding up the complexity associated with constructing a
BFS tree and a random circulation, the fc-cast, and the
converge-cast, we obtain $O(\Diam+\Diam+\Diam+\Diam)$ time and
$O(E+E+\min\{E\Diam, V^2\}+V\Diam) = O(\min\{E\Diam, V^2\})$
messages, as claimed.\end{proof}

\ignore{\subsubsection{Proofs of Claims}\label{sec:phish}
\begin{proof}[Proof of \prettyref{clm:phish1}]
If $e$ is not in any cut pair, then by \eqref{eq:ass2}, $\phi(e)
\neq \phi(f)$ for every $f \neq e$, so $\phi(e)$ occurs once in
$\Phi_e$.

If $e$ is in a cut pair $\{e, f\}$, \prettyref{clm:responsible}
implies that $\{e, f\} \in C_e$ (because no fundamental cycle but
$C_e$ contains $e$). By \eqref{eq:ass2} $\phi(e) = \phi(f)$. Since
$\{e, f\} \subset C_e$, $\phi(e)$ occurs at least twice in
$\Phi_e$.
\end{proof}

\begin{proof}[Proof of \prettyref{clm:phish2}]
If
$\{v, p(v)\}$ does not lie in any cut pair,
then by \eqref{eq:ass2}, $\phi(\{v, p(v)\})$ cannot occur multiple times
in any $\Phi_e$. Hence, $\w[u, v] = 0$ for all $u$.

If $\{v, p(v)\}$ lies in some cut pair, then by
\prettyref{clm:responsible} there is some non-tree edge $e$ so that
$C_e$ contains the cut class of $\{v, p(v)\}$. Let $u$ be either
endpoint of $e$; since $\{v, p(v)\} \in C_e$, $u$ is indeed a
descendant of $v$. Due to \eqref{eq:ass2}, $\w[u, v] > 0$.
\end{proof}}

\section{Computing $\{2,3\}$-Edge-Connected Components}\label{sec:cc}
Let $E_{C}$ denote the set of all cut edges, and $E_{CP}$ denote the
set of all edges in any cut pair.

\begin{defn}
The \emph{$2$-edge-connected components} are the connected
components of $G \bs E_C$. The \emph{$3$-edge-connected components}
are the connected components of $G \bs (E_{CP} \cup E_C)$.
\end{defn}

In the sequential model, connected components of a graph can be
computed in linear time. Hence we immediately see that our
linear-time sequential cut edge and cut pair algorithms yield
linear-time algorithms for 2- and 3-edge-connected components.

In the distributed model, we first discuss 2-edge-connected
components. Let $T$ denote a spanning tree and $r$ its root. The
desired representation is for each vertex $v$ to store a label
$\tau(v)$ so that $\tau(u)=\tau(v)$ iff $u, v$ are in the same
2-edge-connected component. Observe that $E_C \subset E(T)$, since
if $e \not\in T$, then $G \bs e \supset T$ is connected.
Furthermore, the following holds.
\begin{clm}
If $u, v$ are in the same $2$-edge-connected component, there are no
cut edges on the unique $u$-$v$ path in $T$.
\end{clm}
\begin{proof}
Suppose such a cut edge $e = \{u', v'\}$ exists, where $u'$ is the
end of $e$ closer to $u$ along the $u$-$v$ path in $T$. Then in $G
\bs \{e\}$, the remainder of the tree path connects $u$ to $u'$ and
$v$ to $v'$. Since $u, v$ are in the same 2-edge-connected
component, $u$ and $v$ are connected in $G \bs \{e\}$. Thus $u'$ and
$v'$ are connected in $G \bs \{e\}$, contradicting the fact that $e
= \{u', v'\}$ is a cut edge of $G$.
\end{proof}

\begin{cor}\label{cor:2ecc-struct}
$T \bs E_C$ is a spanning forest of the
$2$-edge-connected components.
\end{cor}

In particular, for each 2-edge-connected component $H$, there is a
subtree $T_H$ of $T \bs E_C$ spanning $H$. The idea is to label the
vertices of $H$ by the ID of the root of $T_H$.

\begin{thm}\label{thm:2ecc}
There is a distributed algorithm to compute all $2$-edge-connected
components with probability at least $1-1/V$ in $O(\Diam)$ time and using
$O(E)$ messages.
\end{thm}
\begin{proof}
Note for a vertex $v$, where $H$ denotes its 2-edge-connected
component, $v$ is the root of $T_H$ if and only if either $v$ is the root $r$ of $T$,
or $\{v, p(v)\}$ is a cut edge. Otherwise, $v$ and $p(v)$ are in the
same 2-edge-connected component.

First we compute the cut edges, using \prettyref{thm:dce}. Vertex
$r$ sets $\tau(r)$ equal to its ID. In the following $h(T)$ rounds,
for $i=1$ to $h(T)$, for all level-$i$ tree edges $\{v, p(v)\}$ in
parallel, vertex $p(v)$ sends $\tau(p(v))$ to $v$. Upon receiving
this message, $v$ sets $\tau(v) := ID(v)$ if $\{v, p(v)\}$ is a cut
edge, and $\tau(v) := \tau(p(v))$ otherwise.

The labeling takes $O(h(T))$ time and $|V|-1$ messages, and the
result follows.
\end{proof}

Now we discuss 3-edge-connected components. In the distributed
model, we can represent a subgraph $(V, F)$ of $(V, E)$ by using a
local boolean variable for each edge. For this representation,
\cite{Thur97} gave a distributed connected components
algorithm in $O(\Diam+\sqrt{V}\log^* V)$ time, using an MST subroutine in which the weight of edge $e$ is 1 for
$e \not\in F$ and 0 for $e \in F$.
Hence we have the
following corollary to our cut pair algorithm, \prettyref{thm:dce}.
\begin{cor}\label{cor:dce}
There is a distributed algorithm to compute all 3-edge-connected
components with probability at least $1-1/V$ in $O(\Diam+\sqrt{V}\log^* V)$
time and using $O(E(\Diam+\sqrt{V}\log^* V))$ messages.
\end{cor}

\section{Las Vegas Distributed Implementation} \label{sec:lasvegas}
In this section we describe how to turn our Monte Carlo distributed
algorithms into Las Vegas algorithms, by giving a
\emph{verifier} for each one. Given the output of the Monte Carlo algorithm,
the verifier determines whether the output is correct or not; we re-run the Monte Carlo algorithm until the output is verified correct. For
each of our verifiers, the
time complexity is no more than the time complexity
of the corresponding Monte Carlo algorithm; this fact and the fact
that our algorithms work with high probability together imply that
the resulting Las Vegas algorithms have the same asymptotic
complexity as the Monte Carlo ones. See \cite[\S 1.2]{randalgs} for
more details.

Here is a high-level description of the three verifiers. The cut edge verifier works by labeling vertices according to their 2-edge-connected component; the cut vertex verifier works by labeling edges according to their \emph{blocks}; the cut pair verifier works by exploiting relations between cut classes and fundamental cycles. All three of the verifiers rely on the fact that our Monte Carlo algorithms have one-sided error.

\subsection{Verifier for Cut Edges}\label{lv:cut edge}
Recall that \prettyref{alg:edge} always outputs all cut edges, but may erroneously
output some non-cut edges. Observe that a non-tree edge cannot be a cut edge; so we may assume the Monte Carlo algorithm outputs a set $E'_C$ such that $E(T) \supseteq E'_C \supseteq E_C$, by having the verifier reject any output containing a non-tree edge. Here is the key idea: we compute the connected components of $T \bs E'_C$. We only need to show how to determine if $E'_C \bs E_C$ is nonempty; this can be done using the following proposition and its converse, which follows.

\begin{prop}\label{prop:vce}If $E'_C \bs E_C$ is nonempty, there is a non-tree edge joining vertices in different connected components of $T \bs E'_C$.
\end{prop}
\begin{proof}
Let $e$ be any element of $E'_C \bs E_C$. Since $e$ is not a cut edge, there is another edge $f \in E$ connecting the two connected components of $T \bs e.$ The endpoints of $f$ lie in different connected components of $T \bs E'_C$.
\end{proof}
\begin{prop}\label{prop:covce}
If $E'_C \bs E_C$ is empty, then the connected components of $T \bs E'_C$ are the 2-edge-connected components, and every non-tree edge has its endpoints in the same connected component of $T \bs E'_C$.
\end{prop}
\begin{proof}
\prettyref{cor:2ecc-struct} guarantees that the connected components of $T \bs E'_C$ are the 2-edge-connected components of $G$. Since each non-tree edge lies in at least one cycle (e.g.\ its fundamental cycle with $T$) its endpoints lie in the same 2-edge-connected component.
\end{proof}
\begin{thm}\label{thm:lvedge}
There is a Las Vegas distributed algorithm to compute all cut edges in $O(\Diam)$ time and using $O(E)$ messages, in expectation.
\end{thm}
\begin{proof}
We run the $O(\Diam)$-time, $O(E)$-message Monte Carlo cut edge algorithm from \prettyref{thm:dce}, and as remarked earlier, we know its output $E'_C$ satisfies $E'_C \supseteq E_C$. Then we run the following verifier, terminating if it accepts, and restarting from scratch (i.e., re-running the Monte Carlo algorithm) as long as it rejects.

If $E'_C$ contains a non-tree edge, we reject. Otherwise (if $E'_C \subset E(T)$) we compute the connected components of $E(T) \bs E'_C$ using an implementation like that in the proof of \prettyref{thm:2ecc}, which takes $O(V)$ messages and $O(\Diam)$ time. If any non-tree edge has both endpoints in different components we reject, otherwise the verifier accepts; this can be checked in unit time and $O(E)$ messages. It follows from Propositions \ref{prop:vce} and \ref{prop:covce} that the verifier accepts if and only if $E'_C = E_C$. Since the probability of acceptance is $\Omega(1)$, the expected time complexity is $O(\Diam+\Diam+1)$ and the expected message complexity is $O(E+V+E)$.
\end{proof}

\subsection{Verifier for Cut Pairs}\label{sec:vcp}
As in \prettyref{sec:cutpairimpl} we assume without loss of generality in this section that $G$ is 2-edge-connected.

Consider the output of our Monte Carlo cut pair algorithm, \prettyref{alg:pair}. The sense in which its output is one-sided is that every cut class is a subset of one of its output classes; the verifier must ensure that no cut class is ``too big." To explain our approach, we define a notion of ``wanting." Recall $\Phi_e$, the multiset $\{\phi(f) \mid f \in C_e\}$ defined in \prettyref{sec:cutpairimpl}; if the value $x$ appears more than once in $\Phi_e$, say that $e$ \emph{wants} the set $\{f \in C_e \mid \phi(f) = x\}$. With high probability, the wanted sets are precisely the cut classes. First, our verifier checks that whenever an edge lies in two wanted sets, those sets are the same; second, we use the following proposition to verify that no wanted set is ``too big."
\begin{prop} \label{prop:outsuf}
Let $T$ be any spanning tree and $e, f$ be edges that are not cut edges. If $\{e, f\}$ is not a cut pair, then some fundamental cycle
of $T$ contains exactly one of $e$ and $f.$
\end{prop}
\begin{proof}
We prove the contrapositive; hence we assume that the characteristic vector of $\{e, f\}$ has even dot product with every fundamental cycle. By \prettyref{prop:coolio}(c) the fundamental cycles form a basis of the cycle space; so $\{e, f\}$ is orthogonal to the cycle space, and by \prettyref{prop:coolio}(a), lies in the cut space. Thus $\{e, f\}$ is an induced edge cut, and so (by \prettyref{prop:foo}) a cut pair.
\end{proof}

In order to apply \prettyref{prop:outsuf}, we count the size of all wanted sets, since then each non-tree edge can determine if its fundamental cycle is ``missing" some members.
Our strategy uses a modified {\sc Converge-Cast} (\prettyref{prop:con-cast}) where we interpret $\max$ as lexicographic comparison of data. We need to give each edge a distinct $O(\log V)$-bit name, e.g.\ by concatenating the IDs of its endpoints.
When $e$ wants $S$, it sends the ordered pair $(e, |S|)$ towards all of $S.$ (Concretely, for each tree edge $\{v, p(v)\}$ in $S$, this data is sent to $v$.) If two pairs $(e, k)$ and $(e', k')$ such that $k \neq k'$ are sent to the same location, the verifier rejects. Otherwise, each tree edge takes the label $(e, k)$ where $e$ is the lexicographically-maximal edge that wants it.
We run another fc-cast with the new labels; then each non-tree edge $f$ checks, for each distinct label $(e, k)$ occurring in $C_f$, that there are exactly $k$ edges in $C_f$ with label $(e, k)$. The complexity of the verifier is dominated by the fc-cast, and we thereby obtain the following theorem.
\begin{thm}\label{thm:lvdcpair}
There is a Las Vegas distributed algorithm to compute all cut classes in $O(\Diam)$ time and using $O(\min\{E\cdot
\Diam, V^2\})$ messages, in expectation.
\end{thm}

\subsection{Verifier for Cut Vertices and Blocks}
For edges $e,
f$ in $E(G)$, define $e \sim f$ if either $e=f,$ or $e \neq f$ and there is a
cycle that contains both $e$ and $f.$ It is well-known that $\sim$ is an equivalence relation on $E$; its equivalence classes are called the \emph{blocks} of $G$. A vertex is a cut vertex iff it is incident to more than one block. The overall strategy is to try to label the edges according to the blocks, and then check via a generating relation that our labeling is correct.

The strategy for this verifier is more involved than for the other two, and a high-level description is as follows.
Given two equivalence relations $R$ and $R'$ on the same set, we say that \emph{$R$ refines $R'$} if every equivalence class of $R$ is a subset of some equivalence class of $R'$. Note that $R$ refines $R'$ and $R'$ refines $R$ if and only if $R = R'$.
We use the notion of \emph{local blocks}:
\begin{defn}
The \emph{local blocks at $v$,} denoted $\sim_v$, is an equivalence relation on $\delta(v)$ obtained by restricting $\sim$ to $\delta(v):$ namely we write $e \sim_v f$ iff $e, f \in \delta(v)$ and $e \sim f$.\end{defn}
An analogue of \prettyref{clm:apa} will show that with high probability, the linear dependencies amongst columns of $M^{[v]}$ correspond to the local blocks at $v$.
We hence compute equivalence relations $\sim'_v$ on $\delta(v)$, for each $v$, with the following properties:
\begin{itemize}
\item $\sim'_v$ always refines $\sim_v$
\item we can collect the local relations $\sim'_v$ into a global equivalence relation $\sim'$ on $E$
\item $\sim'$ always refines $\sim$
\item with high probability, $\sim'_v = \sim_v$ for all $v$
\item if $\sim'_v = \sim_v$ for all $v$, then $\sim' = \sim$
\end{itemize}
Finally, we need to check whether $\sim' = \sim$. To perform this check, we adapt
 an approach from work of \cite{tarjan-parallel} and \cite{Thur97}, exemplified in the following proposition, which we will prove in \prettyref{sec:zion}.
\begin{prop}\label{prop:sim0}\label{prop:aaa}
In $O(\Diam)$ time and $O(E)$ messages we can compute a relation $\sim_0$ on $E$ so that (1) whenever $e \sim_0 f$, $e$ and $f$ meet at a vertex, and (2) the symmetric reflexive transitive closure of $\sim_0$ is $\sim$.
\end{prop}
Some logical manipulation shows that $\sim \textrm{ refines } \sim'$ if and only if
$$\forall v : (\forall u, w \textrm{ adjacent to } v: \{u, v\} \sim_0 \{v, w\} \Rightarrow \{u, v\} \sim' \{v, w\})$$
and as a result, local checks complete the verification. We now give the details.

\subsubsection{Computing $\sim'_v$}\label{sec:iron}
What do the local blocks look like? It is not hard to see that the local blocks at $v$ correspond to the connected components of $G \bs v$, in the sense that $\{u, v\} \sim_v \{w, v\}$ if and only if $u$ and $w$ are connected in $G \bs v$. It is also straightforward to see that $F \subset \delta(v)$ is an induced edge cut if and only if $F$ is a disjoint union of equivalence classes of $\sim_v$. We take $b = \lceil\Delta + 2 \log_2 V\rceil$ and just as in \prettyref{clm:apa}, with probability $1-O(1/V^2)$, the following ``good" case holds: the minimal sets of linearly dependent columns of $M^{[v]}$ correspond to the parts of $\sim_v$. (Notice that $C$ is a minimal set of linearly dependent columns iff $C$'s sum is the zero vector and no subset of $C$ adds to the zero vector.) This leads to a simple idea, but we need to use some finesse in order that the $\sim'_v$ we compute from $M^{[v]}$ always refines $\sim_v$.

Our starting point is to compute an arbitrary partition $\pi$ of the columns of $M^{[v]}$ into minimal zero-sum sets (such a partition exists because the sum of all columns is zero). It is possible that such a partition does not refine $\sim'_v$; so we need to check an additional property of $\pi$, namely that each pair of parts of $\pi$ has mutually orthogonal span. (If this property does not hold, the verifier rejects and we re-start the Monte Carlo algorithm.) This property ensures that the only zero-sum sets of columns are unions of parts of $\pi$, which in turn shows that $\sim_v$ refines $\pi$. (Moreover, this property holds in the ``good" case.) So we obtain $\sim'_v$ from $\pi$ by replacing each column by its index in $\delta(v)$.

\subsubsection{Computing $\sim'$ from $\sim'_v$}\label{sec:lion}
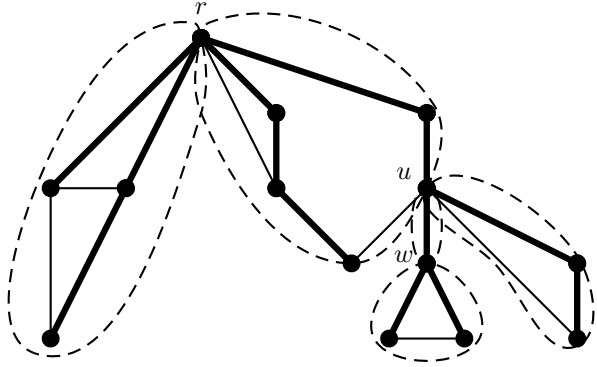
\begin{figure}[t]
\begin{center} \leavevmode
\begin{pspicture}(0,0)(7, 4.2)
  \pnode(2,4){a}\rput(2,4.4){$r$}
  \pnode(0,2){b}
  \pnode(1,2){c}
  \pnode(3,3){d}
  \pnode(5,3){e}
  \pnode(0,0){f}
  \pnode(3,2){g}
  \pnode(5,2){h}\rput(4.7,2.2){$u$}
  \pnode(4,1){i}
  \pnode(7,1){j}
  \pnode(5,1){k}\rput(4.7,1.1){$w$}
  \pnode(7,0){l}
  \pnode(4.5,0){m}
  \pnode(5.5,0){n}
  \psset{linewidth=2.4pt}
  \ncline{*-*}{a}{b}
  \ncline{*-*}{a}{c}
  \ncline{*-*}{a}{d}
  \ncline{*-*}{a}{e}
  \ncline{*-*}{c}{f}
  \ncline{*-*}{d}{g}
  \ncline{*-*}{e}{h}
  \ncline{*-*}{g}{i}
  \ncline{*-*}{h}{j}
  \ncline{*-*}{h}{k}
  \ncline{*-*}{j}{l}
  \ncline{*-*}{k}{m}
  \ncline{*-*}{k}{n}
  \psset{linewidth=0.8pt}
  \ncline{*-*}{b}{c}
  \ncline{*-*}{b}{f}
  \ncline{*-*}{a}{g}
  \ncline{*-*}{h}{i}
  \ncline{*-*}{h}{l}
  \ncline{*-*}{m}{n}
  \psset{linestyle=dashed}
  \psccurve(-0.2,-0.2)(2,3)(2,4)(1.9,4.2)(-0.2,2)
  \psccurve(2,3)(2,4)(2.1,4.2)(5.1,3.1)(5,2)(4,1)
  \psccurve(5,1)(5.2,1.5)(5,2)(4.8,1.5)
  \psccurve(5,1)(5.7,0)(4.3,0)
  \psccurve(5,2)(7,1.1)(7,-0.1)(5.9,1)
\end{pspicture}
\end{center}
\caption{The interaction between a spanning tree and the blocks of a graph. Thick lines are tree edges, thin lines are non-tree edges, and the dashed regions indicate the five blocks of the graph.} \label{fig:tblock}
\end{figure}

For the rest of the section we consider the spanning tree $T$ upon which our algorithms operate as fixed; hence when we say ``fundamental cycle of $e$" we mean with respect to $T$. We assume $T$ is rooted at the leader vertex $r$ and we let $p(v)$ denote the
parent of $v$ in $T.$ In collecting the local relations into a global relation, it is instructive to consider the interaction between $T$ and the blocks of the graph; Figure \ref{fig:tblock} gives an illustration. It is not hard to argue that the intersection of $T$ with any given block $B$ is a subtree of $T$; we define the \emph{root} $r(B)$ of the block to be the root of this subtree. For example, in Figure \ref{fig:tblock}, $r$ and $u$ are each the root of two blocks, and $w$ is the root of one block. In general, the blocks for which $v$ is the root correspond to the equivalence classes of $\sim_v$ not containing $\{v, p(v)\}$ (if $v = r$, all equivalence classes of $\sim_v$).

We now define $\sim'$. For computational purposes, assign each equivalence class $X$ of $\sim_v$ a number $i_v(X)$, using the numbers $1, 2, \dotsc$ for each $v$. Then assign each block $B$ the label $(r(B), i_{r(B)}(X))$ where the equivalence class $X$ is the intersection of $\delta(r(B))$ with $B$. At a high level, to compute $\sim$ from $\sim_v$, within in each block, we broadcast its label starting from the block's root. Now given $\sim'_v$ instead of $\sim_v$, we can mimic this strategy so as to compute a global relation $\sim'$. We give pseudocode in \prettyref{alg:blabel}; the phrase ``$v$ sets directed label $(v, u)$ to $\ell$" means that $v$ stores $\ell$ as the label of $\{v, u\}$ and notifies $u$ of this fact with a message.

\begin{algorithm}[ht]
\caption{Given local relations $\sim'_v,$ compute a global relation $\sim'.$}\label{alg:blabel}
\begin{algorithmic}[1]
\State {\bf at} each vertex $v$, number the equivalence classes of $\sim'_v$ by $1, 2, \dotsc$
\State {\bf at} each vertex $v$, {\bf for} each equivalence class $X$ of $\sim'_v$ not containing $\{v, p(v)\}$, {\bf for} each $\{v, u\} \in X$, set directed label $(v, u)$ to $(v, i_v(X))$
\State {\bf when} vertex $w$ sets directed label $(w, v)$ to $\ell$, {\bf if} the label of $(v, w)$ exists and is not equal to $\ell$ then FAIL, {\bf else if} directed label $(v, w)$ is unassigned, {\bf for} each $\{v, u\} \sim'_v \{v, w\}$, set directed label $(v, u)$ to $\ell$
\State take the edge labels to identify the equivalence classes of $\sim'$
\end{algorithmic}
\end{algorithm}

Any pair of $\sim'$-wise related edges are connected by a path of edges related pairwise by local $\sim'_v$ relations; since $\sim'_v$ refines $\sim_v$ which is a restriction of $\sim$, we see that $\sim'$ refines $\sim$.
When $\sim'_v = \sim_v$ for all $v$, the preceding discussion implies that $\sim' = \sim$.
The message complexity of \prettyref{alg:blabel} is $O(E)$. When $\sim'_v = \sim_v$ for all $v$, the time complexity is $\Diam$ rounds; if more rounds than this elapse we restart the Las Vegas algorithm.

\subsubsection{The Generating Relation $\sim_0$}\label{sec:zion}
In order to define $\sim_0$ we need a few preliminaries. Let $pre(v)$ denote a preordering of $T$ starting from the root, and for each vertex $v$, let $desc(v)$ denote the number of descendants of $v$. Thus the set of descendants of $v$ is the set of vertices with preorder labels in $\{pre(v), \dotsc, pre(v)+desc(v)-1\}.$
The \emph{subtree-neighbourhood} of $v$ is defined to be $v$'s descendants, in addition to every other vertex that is adjacent to a
descendant of $v$ via a non-tree edge. For each vertex $v$ let
the values $low(v)$ and $high(v)$ denote the minimum and maximum
preorder label in the subtree-neighbourhood of $v.$ Tarjan \cite{tarjan74} introduced these $low$ and $high$ functions; they have been used in several biconnectivity algorithms \cite{tarjan-parallel,Thur97}.

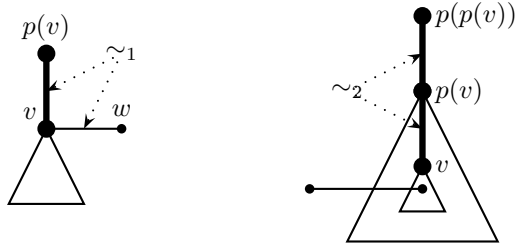
\begin{figure}[t]
\begin{center} \leavevmode
\begin{pspicture}(0,0.4)(10,3)
  \psset{arrowsize=4pt}
  \pnode(2,2.5){lpv}\uput[90](2,2.5){$p(v)$}
  \pnode(2,1.5){lv}\uput[135](2,1.5){$v$}
  \pnode(3,1.5){lw}\uput[90](3,1.5){$w$}
  \pspolygon(2,1.5)(2.5,0.5)(1.5,0.5)
  \ncline[linewidth=2.4pt]{*-*}{lpv}{lv}\mput{\rnode{le}{}}
  \ncline[linewidth=0.8pt]{*-*}{lv}{lw}\mput{\rnode{lf}{}}
  \pnode(3,2.5){lh}\rput(3,2.5){$\sim_1$}
  \ncline[linestyle=dotted]{->}{lh}{le}
  \ncline[linestyle=dotted]{->}{lh}{lf}
  \pnode(7,3){rppv}\uput[0](7,3){$p(p(v))$}
  \pnode(7,2){rpv}\uput[0](7,2){$p(v)$}
  \pnode(7,1){rv}\uput[0](7,1){$v$}
  \pnode(7,0.7){rs}
  \pnode(5.5,0.7){rt}
  \pspolygon(7,2)(6,0)(8,0)
  \pspolygon(7,1)(7.3,0.4)(6.7,0.4)
  \ncline[linewidth=2.4pt]{*-*}{rppv}{rpv}\mput{\rnode{re}{}}
  \ncline[linewidth=2.4pt]{*-*}{rpv}{rv}\mput{\rnode{rf}{}}
  \ncline[linewidth=0.8pt]{*-*}{rs}{rt}
  \pnode(6,2){rh}\rput(6,2){$\sim_2$}
  \ncline[linestyle=dotted]{->}{rh}{re}
  \ncline[linestyle=dotted]{->}{rh}{rf}
\end{pspicture}
\end{center}
\caption{Schematic illustrations of the relations $\sim_1$ (left) and $\sim_2$ (right). Thick edges are tree edges,
thin edges are non-tree edges, and triangles depict sets of descendants.
Dotted arrows indicate pairs of edges related by $\sim_i$.} \label{fig:sim0}
\end{figure}

\begin{defn}
\label{defn:nice}
The relation $\{w, v\}\sim_1\{v,p(v)\}$ holds if and
only if $\{w, v\} \not\in T$ and either $pre(w) < pre(v)$ or $pre(w)
\geq pre(v)+desc(v)$ (i.e., if $w$ is not a descendant of $v$). The relation $\{v, p(v)\} \sim_2
\{p(v),p(p(v))\}$ holds if and only if either $low(v) < pre(p(v))$
or $high(v) \geq pre(p(v))+desc(p(v))$ (i.e., if the subtree-neighbourhood of $v$ is not contained in the descendants of $p(v)$). Define $\sim_0$ to be the union of $\sim_1$ and $\sim_2$.
\end{defn}
We illustrate these relations in Figure \ref{fig:sim0}. Earlier work \cite{tarjan-parallel,Thur97} uses a different generating relation for $\sim$; ours is simpler and also has the crucial property that every two edges related by $\sim_0$ have a common vertex.

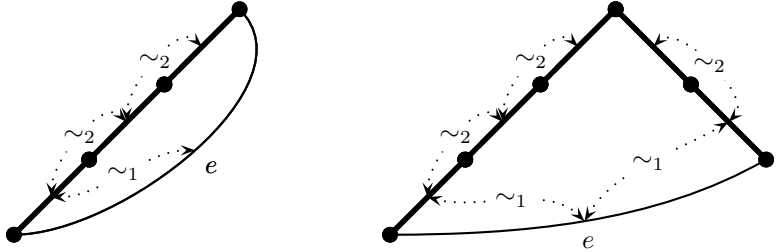
\begin{figure}[t]
\begin{center} \leavevmode
\begin{pspicture}(0,0)(10,3)
  \psset{arrowsize=4pt}
  \pnode(3,3){a}
  \pnode(2,2){b}
  \pnode(1,1){c}
  \pnode(0,0){d}
  \ncline[linewidth=2pt]{*-*}{a}{b}\mput{\rnode{e}{}}
  \ncline[linewidth=2pt]{*-*}{b}{c}\mput{\rnode{f}{}}
  \ncline[linewidth=2pt]{*-*}{c}{d}\mput{\rnode{g}{}}
  \nccurve[angleA=-45,angleB=0]{a}{d}\lput(0.5){\rnode{h}{}}\Aput{$e$}
  \nccurve[linestyle=dotted,angleA=135,angleB=90]{<->}{e}{f}\mput*{$\sim_2$}
  \nccurve[linestyle=dotted,angleA=135,angleB=90]{<->}{f}{g}\mput*{$\sim_2$}
  \nccurve[linestyle=dotted,angleA=0,angleB=180]{<->}{g}{h}\mput*{$\sim_1$}
  \ncline[linewidth=2pt]{*-*}{a}{b}\mput{\rnode{e}{}}
  \ncline[linewidth=2pt]{*-*}{b}{c}\mput{\rnode{f}{}}
  \ncline[linewidth=2pt]{*-*}{c}{d}\mput{\rnode{g}{}}
  \nccurve[angleA=-45,angleB=0]{a}{d}\lput(0.5){\rnode{h}{}}\Aput{$e$}
  \pnode(8,3){A}
  \pnode(7,2){B}
  \pnode(6,1){C}
  \pnode(5,0){D}
  \pnode(9,2){E}
  \pnode(10,1){F}
  \ncline[linewidth=2pt]{*-*}{A}{B}\mput{\rnode{G}{}}
  \ncline[linewidth=2pt]{*-*}{B}{C}\mput{\rnode{H}{}}
  \ncline[linewidth=2pt]{*-*}{C}{D}\mput{\rnode{I}{}}
  \ncline[linewidth=2pt]{*-*}{A}{E}\mput{\rnode{L}{}}
  \ncline[linewidth=2pt]{*-*}{E}{F}\mput{\rnode{K}{}}
  \nccurve[angleA=0,angleB=210]{D}{F}\lput(0.5){\rnode{J}{}}\Bput{$e$}
  \nccurve[linestyle=dotted,angleA=135,angleB=90]{<->}{G}{H}\mput*{$\sim_2$}
  \nccurve[linestyle=dotted,angleA=135,angleB=90]{<->}{H}{I}\mput*{$\sim_2$}
  \nccurve[linestyle=dotted,angleA=45,angleB=45]{<->}{K}{L}\mput*{$\sim_2$}
  \nccurve[linestyle=dotted,angleA=-15,angleB=135]{<->}{I}{J}\mput*{$\sim_1$}
  \nccurve[linestyle=dotted,angleA=60,angleB=210]{<->}{J}{K}\mput*{$\sim_1$}
  \ncline[linewidth=2pt]{*-*}{A}{B}\mput{\rnode{G}{}}
  \ncline[linewidth=2pt]{*-*}{B}{C}\mput{\rnode{H}{}}
  \ncline[linewidth=2pt]{*-*}{C}{D}\mput{\rnode{I}{}}
  \ncline[linewidth=2pt]{*-*}{A}{E}\mput{\rnode{L}{}}
  \ncline[linewidth=2pt]{*-*}{E}{F}\mput{\rnode{K}{}}
\end{pspicture}
\end{center}
\caption{The fundamental cycle $C_e$ in the proof of
\prettyref{prop:aaa}. Edges of $T$ are thick lines and $e$ is labeled.
The left diagram shows the case that one of $e$'s endpoints is a $T$-descendant of
the other, while the right diagram shows the case that $e$'s endpoints are unrelated. Dotted arrows indicate
pairs of edges related by $\sim_i$.} \label{fig:zz}
\end{figure}

From now on, given a relation $R$, let $R^*$ denote the equivalence relation obtained by taking the reflexive symmetric transitive closure of $R$. We now prove the key property of $\sim_0$.
\begin{proof}[Proof of $\sim^*_0 = \sim$ (\prettyref{prop:sim0})]
First, we argue that $\sim_0$ refines $\sim$; for this it suffices to show that when $e \sim_i f$ for $i \in \{1, 2\}$, $e$ and $f$ lie in the same block. If $\{w, v\} \sim_1 \{v,p(v)\}$, the fundamental cycle of $\{v, w\}$ contains $\{v, p(v)\}$, so $\{v, w\} \sim \{v, p(v)\}$ as needed. If $\{v, p(v)\} \sim_2 \{p(v), p(p(v))\}$ then there is edge from a descendant of $v$ to a non-descendant of $p(v)$; the fundamental cycle of this edge contains both $\{v, p(v)\}$ and $\{p(v), p(p(v))\}$, as needed.

Second, we must show that $\sim$ refines $\sim^*_0$.
Define $e \sim_{FC} f$ if $e$ and $f$ lie on a common fundamental cycle. Tarjan \& Vishkin \cite[Thm.~1]{tarjan-parallel} show that $\sim_{FC}^* = \sim.$ So it suffices to show that when $e \sim_{FC} f$, $e \sim_0^* f$ holds. In other words, we need to show that each fundamental cycle lies in a single equivalence class of $\sim_0^*$. We provide a pictorial argument of this fact in Figure \ref{fig:zz}.\end{proof}

We now recap the distributed implementation of our cut vertex verifier.
\begin{thm}\label{thm:lvdcv}
There is a Las Vegas distributed algorithm to compute all cut vertices in $O(\Diam + \Delta / \log V)$ time and using
$O(E (1+ \Delta / \log V ))$ messages, in expectation.
\end{thm}
\begin{proof}
We compute a random $b$-bit circulation for $b = \lceil \Delta + 2\log_2 V \rceil$ and use the resulting values to compute local relations $\sim'_v$. (As mentioned in \prettyref{sec:lion} the verifier may reject at this stage.) We then combine this information into a global labeling $\sim'$ of edges (and again, the verifier may reject at this stage).

There is a straightforward distributed protocol to compute $pre(v), desc(v), low(v)$ and $high(v)$ at each $v$ in $O(h(T))=O(\Diam)$ time and using $O(E)$ messages; see e.g.\ \cite{dp-thesis,Thur97}. After this, each vertex sends these four values to all of its neighbours, with communication taking place along all edges in parallel; this takes $O(1)$ time and $O(E)$ messages.

At this point, for each pair $e, f$ of edges that are related by $\sim_0$, their common endpoint $v$ checks that $e \sim' f$ holds. If there is a violation at any vertex, the verifier rejects, and if not, the verifier accepts. The labels $\sim'$ give the blocks; vertex $v$ is a cut vertex iff at least two blocks meet at $v$.

Computing $\phi$ dominates the time and message complexity; each other step takes $O(\Diam)$ time and $O(E)$ messages. Noting that the verifier accepts each time with probability at least $1-1/V$, \prettyref{thm:lvdcv} follows.
\end{proof}

\section{Lower Bounds on Distributed Time}\label{sec:lowerbounds}
In this section we give precise assumptions under which our distributed cut edge and cut pair algorithms achieve universal
optimality.
Let $r$ denote the unique leader vertex in the graph. A vertex is
\emph{quiescent} in a given round if it does not send any messages
or modify its local memory in that round. We adopt the following
terminology from \cite[\S 3.4 \& Ch.\ 24]{p2000}.

\begin{defn}
A distributed algorithm has \emph{termination detection} if $r$ has
a local boolean variable {\tt done}, initialized to \textsc{false},
so that {\tt done} is set to \textsc{true} exactly once, in the last
round of the algorithm. A distributed algorithm has \emph{a single
initiator} if, except for $r$, every vertex is quiescent until it
receives a message.
\end{defn}

The \emph{state} of a vertex means the contents of its memory. We omit the straightforward inductive proof of the following standard proposition.
\begin{prop}\label{prop:speed} Let two graphs both contain a vertex $v$ and have the same graph topology and node IDs in the distance-$d$ neighbourhood of $v$. If the same deterministic distributed algorithm is run on both graphs, the state of $v$ is the same in both instances for the first $d-1$ rounds. For a randomized algorithm, the distribution over states of $v$ is the same.
\end{prop}

\begin{figure}[t]
\begin{center} \leavevmode
\begin{pspicture}(0,0)(6,2.3)
  \psellipse[fillstyle=vlines](1,1)(1,1)
  \rput*(1,1){$G$}
  \psline[arrows=*-*](2,1)(2.3,1.7)
  \psline[arrows=*-*](2.3,1.7)(3,2)
  \psline[arrows=*-*](3,2)(3.7,1.7)
  \rput*(4,1.1){$\vdots$}
  \psline[arrows=*-*](2,1)(2.3,0.3)
  \psline[arrows=*-*](2.3,0.3)(3,0)
  \psline[arrows=*-*](3,0)(3.7,0.3)
  \psline[arrows=-](3.7,0.3)(3.85,0.65)
  \psline[arrows=-](3.7,1.7)(3.85,1.35)
  \uput[0](2,1){$v$}
  \uput[90](2.3,1.7){$v_1$}
  \uput[90](3,2){$v_2$}
  \uput[90](3.7,1.7){$v_3$}
    \uput[-90](2.3,0.3){$v_{\ell-1}$}
  \uput[-90](3,0){$v_{\ell-2}$}
    \uput[-90](3.7,0.3){$v_{\ell-3}$}
  \psline[arrows=*-*](0,1)(0,1)
  \uput[180](0,1){$r$}
\end{pspicture}
\begin{pspicture}(0,0)(4,2)
  \psellipse[fillstyle=vlines](1,1)(1,1)
  \rput*(1,1){$G$}
  \psline[arrows=*-*](2,1)(2.3,1.7)
  \psline[arrows=*-*](2.3,1.7)(3,2)
  \psline[arrows=*-*](3,2)(3.7,1.7)
  \rput*(4,1.1){$\vdots$}
  \psline[arrows=*-*](2.3,0.3)(3,0)
  \psline[arrows=*-*](3,0)(3.7,0.3)
  \psline[arrows=-](3.7,0.3)(3.85,0.65)
  \psline[arrows=-](3.7,1.7)(3.85,1.35)
  \uput[0](2,1){$v$}
  \uput[90](2.3,1.7){$v_1$}
  \uput[90](3,2){$v_2$}
  \uput[90](3.7,1.7){$v_3$}
    \uput[-90](2.3,0.3){$v_{\ell-1}$}
  \uput[-90](3,0){$v_{\ell-2}$}
    \uput[-90](3.7,0.3){$v_{\ell-3}$}
  \psline[arrows=*-*](0,1)(0,1)
  \uput[180](0,1){$r$}
\end{pspicture}
\end{center}
\caption{Left: the graph $G_c$. Right: the graph
$G_p$.} \label{fig:lb}
\end{figure}
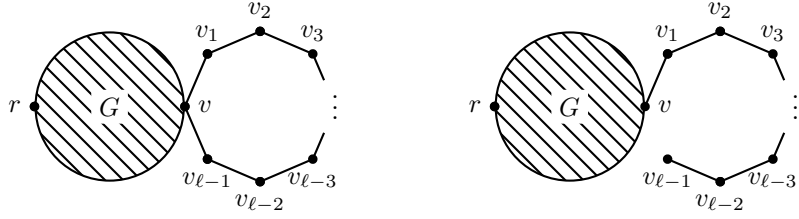

For a graph $G$, a vertex $v \in V(G)$ and an integer $\ell \geq 3$, we now define graphs
$G_c$ and $G_p$ that implicitly depend on $\ell$ and $v$. Specifically, let $G_c$ denote the graph obtained from $G$ by attaching a
$\ell$-edge cycle to $G$ at $v$, and let $G_p$ denote the graph
obtained from $G$ by attaching a $(\ell-1)$-edge path to $G$ at $v$,
as shown in Figure \ref{fig:lb}. Give corresponding vertices $v_i$
in the two graphs the same ID.

\begin{thm}\label{thm:tdlower}
Any deterministic distributed algorithm for finding all cut edges that has
termination detection takes at least $\Diam/2$ rounds on every graph.
\end{thm}
\begin{proof}
Consider for the sake of contradiction a graph $G$ upon which the algorithm terminates in
$t < \Diam/2$ rounds. Let $v$ be any vertex of distance at
least $\Diam/2$ away from $r$, and let $\ell = 2t + 2$. By \prettyref{prop:speed},
the algorithm also sets ${\tt done} := \textsc{true}$ at $r$ on $G^{}_p$ and
$G^{}_c$ in $t$ rounds, so the algorithms terminate then.

Now consider $v_{\ell/2}$; using \prettyref{prop:speed} again, we see that its state is the same at termination in both instances. Since the edges incident to $v_{\ell/2}$ are cut edges in $G^{}_p$ but not in $G^{}_c$, they must have been incorrectly classified at $v_{\ell/2}$ in at least one instance.
\end{proof}

If we assume that the algorithm has a single initiator instead of
assuming termination detection, a similar argument works. We use the
following lemma, whose easy inductive proof is omitted.
\begin{lmma}\label{lmma:speed2}
In an algorithm with a single initiator, every vertex at distance
$t$ from $r$ is quiescent for the first $t$ rounds.
\end{lmma}
\begin{thm}\label{thm:sing-lb}
Any deterministic distributed algorithm for finding all cut edges that has
a single initiator takes at least $\Diam/2$ rounds on every graph.
\end{thm}
\begin{proof}
Suppose the algorithm terminates in $t<\Diam/2$ rounds on a graph $G$.
Let $v$ be any vertex of distance at least $\Diam/2$ away from $r$. Then by \prettyref{prop:speed} the algorithm also terminates in
$t$ rounds on $G^{3,v}_c$ and $G^{3,v}_p$. By
\prettyref{lmma:speed2} vertex $v_1$ is
quiescent during the entire execution of the algorithm on these new graphs; hence the incident edges cannot be correctly classified in both instances.
\end{proof}
For randomized algorithms we have the following lower bound.
\begin{thm}
Any randomized distributed algorithm with error probability less than $1/4$ for finding all cut edges takes at least $\Diam/4$ rounds in expectation, if it has a single initiator or termination confirmation.
\end{thm}
\begin{proof}
We use the same setup as in the proofs of Theorems \ref{thm:tdlower} and \ref{thm:sing-lb}.
Markov's inequality shows that when running the algorithm on $G$, the time of termination $t$ satisfies $\Pr[t \leq \Diam/2] \geq 1/2.$
The distribution on the state of the crucial vertex --- $v_{\ell/2}$ for termination confirmation, $v_1$ for single initiator --- is the same on both $G_c$ and $G_p$ at time $\Diam/2$. So of the $\geq 1/2$ probability mass of termination before $\Diam/2$, either $1/4$ incorrectly classifies edges of $G_c$ as cut edges or edges of $G_p$ as not cut edges.
\end{proof}

The same lower bounds hold for finding
2-edge-connected components and cut pairs, since the new edges of
$G^{}_c$ are in cut pairs, while the new edges of
$G^{}_p$ are not. It is straightforward to verify that our
distributed algorithms can be implemented so as to have a single
initiator and termination detection; then their universal
optimality follows.

If we do not require a single initiator or
termination detection, and if we change our input model to allow
additional parameters of $G$ to be initially known at each node,
\emph{neighbourhood cover} techniques of \cite{Elkin06} can be
synthesized with our techniques to yield even faster algorithms for
certain graph classes. Elkin used these techniques to obtain
distributed MST algorithms faster than $O(\Diam)$ on some graphs.

\section{Parallel Cut Pairs on the EREW PRAM} \label{sec:parallel}
In this section we give a parallel cut pair algorithm of time complexity $O(\log V)$ for the EREW PRAM. Computing the OR of $n$ bits has a lower bound of $\Omega(\log n)$ time in this model; from this an easy combinatorial reduction yields an $\Omega(\log V)$ time lower bound for finding all cut pairs of a graph, so our algorithm is time-optimal. As in \prettyref{sec:cutpairimpl} we assume without loss of generality in this section that $G$ is 2-edge-connected.

We will require several common subroutines. First, we need a Las Vegas randomized spanning forest subroutine taking $O(V+E)$ work and space and $O(\log V)$ time, due to \cite{HZ01}. An \emph{ear decomposition} can be computed in the same randomized complexity using the approaches in \cite{MSV86,MR92} and plugging in the result of \cite{HZ01} for the spanning forest subroutine. \emph{Expression evaluation} of an $n$-node tree can be accomplished deterministically in $O(n)$ work and space and $O(\log n)$ time (e.g.\ see the book \cite[Ch.\ 3]{JaJa92}). We let $T(n), S(n), W(n)$ denote the time, space, work complexity to sort $n$ numbers of length $O(\log n)$ bits; we give references to the best known algorithms for this problem in Section \ref{sec:contributions} (they are deterministic). First, we give our Monte Carlo cut pair algorithm.

\begin{thm}
There is a parallel algorithm to compute all cut pairs with
probability at least $1-1/V$ in $O(\log V + T(E))$ time, $O(E+S(E))$ space, and $O(E+W(E))$ work.
\end{thm}
\begin{proof}
We implement \prettyref{alg:pair} distributively. First, we claim we can implement the subroutine \algob\ distributively to generate a random $O(\log V)$-bit circulation in logarithmic time and linear work; the completion steps (\prettyref{alg:complete}) are accomplished via a call to expression evaluation in which we compute the expression $\phi(e) := \bigoplus_{f \in \delta(v) \bs e} \phi(f)$ for each tree edge $e = \{v, p(v)\}$. We implement \prettyref{line:val-loop} of \prettyref{alg:pair} via a sort.
\end{proof}


\subsection{Las Vegas Cut Pair Algorithm}
The verifier for our parallel cut pair algorithm works by attempting to construct the \emph{2-cactus} of $G$ which acts as a certificate for all of the cut pairs. Our terminology is derived from a more general sort of cactus originally due to \cite{DKL76} 
that acts as a certificate for all minimum edge cuts. Say that $u \equiv v$ in $G$ if the edge-connectivity between $u$ and $v$ is at least 3; it is easy to show (e.g.\ using the max-flow min-cut theorem) that $\equiv$ is an equivalence relation.

We now define how to \emph{contract\footnote{When we ``contract," we may identify \emph{non-adjacent} vertices; this contrasts with the more common meaning of ``contract" in the context of graph minors.} a graph by an equivalence relation}. Contraction may introduce parallel edges and/or loops; for this reason in the rest of the paper, when we say a graph, we mean a \emph{multigraph}, which may have parallel edges and/or loops. Given a graph $G$ and an equivalence relation $R$ on the vertices of $G$, the \emph{contraction} denoted $G/R$ is another (multi)graph.
For each equivalence class $C$ of $R$, $G/R$ has a vertex labelled $C$. For each edge $e$ of $G$, $G/R$ has an edge from the vertex labelled by the equivalence class of $u$ to that of $v$. Since the edges of $G$ correspond bijectively to the edges of $G/R$, we will speak of the graphs as having the same edge set (alternatively, one may think of each edge of $G$ having a distinct label which is inherited by the corresponding edge of $G/R$).

\begin{defn}
The \emph{2-cactus} $\gcac$ of $G$ is $G/\!\equiv$.
\end{defn}

\newcommand{\pnodeput}[2]{\pnode(#1){#2}\ncline[arrows=*-*]{#2}{#2}}

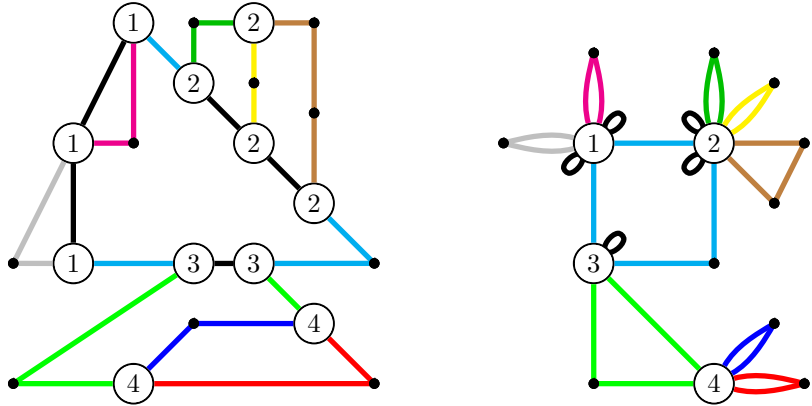
\begin{figure}[t]
\begin{center} \leavevmode
\begin{pspicture}(-0.4,-0.4)(5.2,5.2)
\psset{unit=0.8cm}
  \pnodeput{0,0}{a}
  \cnodeput(2,0){b}{4}
  \pnodeput{6,0}{c}
  \pnodeput{3,1}{d}
  \cnodeput(5,1){e}{4}
  \pnodeput{0,2}{f}
  \cnodeput(1,2){g}{1}
  \cnodeput(3,2){h}{3}
  \cnodeput(4,2){i}{3}
  \pnodeput{6,2}{j}
  \cnodeput(5,3){k}{2}
  \cnodeput(1,4){l}{1}
  \pnodeput{2,4}{m}
  \cnodeput(4,4){n}{2}
  \cnodeput(3,5){p}{2}
  \pnodeput{4,5}{q}
  \cnodeput(2,6){s}{1}
  \pnodeput{3,6}{t}
  \cnodeput(4,6){u}{2}
  \pnodeput{5,6}{v}
  \pnodeput{5,4.5}{o}
  \psset{linewidth=2pt}
  {\psset{linecolor=green}
  \ncline{a}{b}
  \ncline{a}{h}
  \ncline{e}{i}
  }
  {\psset{linecolor=blue}
  \ncline{b}{d}
  \ncline{e}{d}
  }
  {\psset{linecolor=red}
  \ncline{b}{c}
  \ncline{e}{c}
  }
  \ncline{h}{i}
  {\psset{linecolor=cyan}
  \ncline{h}{g}
  \ncline{i}{j}
  \ncline{j}{k}
  \ncline{s}{p}
  }
  {\psset{linecolor=lightgray}
  \ncline{f}{g}
  \ncline{l}{f}
  }
  \ncline{l}{g}
    {\psset{linecolor=magenta}
  \ncline{l}{m}
  \ncline{m}{s}
  }
  \ncline{l}{s}
  \ncline{p}{n}
  \ncline{n}{k}
    {\psset{linecolor=colcite}
  \ncline{p}{t}
  \ncline{t}{u}
  }
    {\psset{linecolor=yellow}
  \ncline{n}{q}
  \ncline{q}{u}
  }
    {\psset{linecolor=brown}
  \ncline{k}{o}
  \ncline{o}{v}
  \ncline{v}{u}
  }
  \psset{linewidth=1pt}
  \pnodeput{0,0}{za}
  \cnodeput*(2,0){zb}{4}
  \pnodeput{6,0}{zc}
  \pnodeput{3,1}{zd}
  \cnodeput*(5,1){ze}{4}
  \pnodeput{0,2}{zf}
  \cnodeput*(1,2){zg}{1}
  \cnodeput*(3,2){zh}{3}
  \cnodeput*(4,2){zi}{3}
  \pnodeput{6,2}{zj}
  \cnodeput*(5,3){zk}{2}
  \cnodeput*(1,4){zl}{1}
  \pnodeput{2,4}{zm}
  \cnodeput*(4,4){zn}{2}
  \cnodeput*(3,5){zp}{2}
  \pnodeput{4,5}{zq}
  \cnodeput*(2,6){zs}{1}
  \pnodeput{3,6}{zt}
  \cnodeput*(4,6){zu}{2}
  \pnodeput{5,6}{zv}
  \pnodeput{5,4.5}{zo}
\end{pspicture}
\begin{pspicture}(-2.4,-0.4)(4,5.2)
\psset{unit=0.8cm}
  \pnodeput{0,0}{a}
  \cnodeput(2,0){b}{4}
  \pnodeput{3.5,0}{c}
  \pnodeput{3,1}{d}
  \cnodeput(0,2){e}{3}
  \pnodeput{2,2}{f}
  \cnodeput(0,4){g}{1}
  \pnodeput{-1.5,4}{h}
  \pnodeput{0,5.5}{i}
  \cnodeput(2,4){j}{2}
  \pnodeput{2,5.5}{k}
  \pnodeput{3,5}{l}
  \pnodeput{3.5,4}{m}
  \pnodeput{3,3}{n}
  \psset{linewidth=2pt}
  {\psset{linecolor=green}
  \ncline{a}{b}
  \ncline{a}{e}
  \ncline{e}{b}
  }
  {\psset{linecolor=blue}
  \nccurve[angleA=-150,angleB=60]{d}{b}
  \nccurve[angleA=-120,angleB=30]{d}{b}
  }
  {\psset{linecolor=red}
  \nccurve[angleA=15,angleB=165]{b}{c}
  \nccurve[angleA=-15,angleB=-165]{b}{c}
  }
    {\psset{linecolor=cyan}
  \ncline{e}{f}
  \ncline{j}{f}
  \ncline{g}{j}
  \ncline{e}{g}
  }
    {\psset{linecolor=lightgray}
  \nccurve[angleA=15,angleB=165]{h}{g}
  \nccurve[angleA=-15,angleB=-165]{h}{g}
  }
    {\psset{linecolor=magenta}
  \nccurve[angleA=75,angleB=-75]{g}{i}
  \nccurve[angleA=105,angleB=-105]{g}{i}
  }
    {\psset{linecolor=colcite}
  \nccurve[angleA=75,angleB=-75]{j}{k}
  \nccurve[angleA=105,angleB=-105]{j}{k}
  }
    {\psset{linecolor=yellow}
  \nccurve[angleA=30,angleB=-120]{j}{l}
  \nccurve[angleA=60,angleB=-150]{j}{l}
  }
    {\psset{linecolor=brown}
  \ncline{j}{m}
  \ncline{n}{m}
  \ncline{j}{n}
  }
  \psccurve(0.2,4.2)(0.5,4.4)(0.4,4.5)(0.2,4.2)
  \psccurve(0.2,2.2)(0.5,2.4)(0.4,2.5)(0.2,2.2)
  \psccurve(1.8,4.2)(1.5,4.4)(1.6,4.5)(1.8,4.2)
  \psccurve(1.8,3.8)(1.5,3.6)(1.6,3.5)(1.8,3.8)
  \psccurve(-0.2,3.8)(-0.5,3.6)(-0.4,3.5)(-0.2,3.8)
  \psset{linewidth=1pt}
  \pnodeput{0,0}{za}
  \cnodeput*(2,0){zb}{4}
  \pnodeput{3.5,0}{zc}
  \pnodeput{3,1}{zd}
  \cnodeput*(0,2){ze}{3}
  \pnodeput{2,2}{zf}
  \cnodeput*(0,4){zg}{1}
  \pnodeput{-1.5,4}{zh}
  \pnodeput{0,5.5}{zi}
  \cnodeput*(2,4){zj}{2}
  \pnodeput{2,5.5}{zk}
  \pnodeput{3,5}{zl}
  \pnodeput{3.5,4}{zm}
  \pnodeput{3,3}{zn}
\end{pspicture}
\end{center}
\caption{Left: a graph. For ease of visualization, every node is labelled by its equivalence class of $\equiv$, except for nodes in singleton classes; every edge is coloured according to its cut class, except for edges in no cut pair, which are black. Right: the 2-cactus $\gcac$ defined equal to $G/\!\equiv$.} \label{fig:2cactus}
\end{figure}

An example of a 2-cactus is given in Figure \ref{fig:2cactus}.
\begin{prop}\label{prop:cpp}
(a) For any equivalence relation $R$ on $V$, every cut pair of $G/R$ is a cut pair of $G$. (b) Every cut pair of $G$ is a cut pair of $\gcac$.
\end{prop}
\begin{proof}
Both results use a common observation. As before let $\delta(Z)$ denote the set of edges with exactly one endpoint in $Z$. Let $X$ be a set of equivalence classes of $R$ (which we may view as a vertex set in $G/R$) and let $\cup X$ be the union of those classes (which is a vertex set in $G$). Then from the definition of contraction it is easy to see the following:
\begin{equation}\label{eq:footz}
\textrm{The edge set $\delta(\cup X)$ of $G$ is the same as the edge set $\delta(X)$ of $G/R$.}\tag{\ddag}\end{equation}

To prove (a), let $\delta(X) = \{e, f\}$ be a cut pair of $G/R$, where $X$ is a set of equivalence classes of $R$ (vertices of $G/R$). Let $\cup X$ be the union of these classes. By \eqref{eq:footz}, in $G$ the edge set $\delta(\cup X)$ is precisely $\{e, f\}$ giving the needed fact that $\{e, f\}$ is a cut pair of $G$.

To prove (b), let $\delta(S) = \{e, f\}$ be a cut pair of $G$. By the (weak) max-flow min-cut theorem $s \nequiv t$ holds for each $s \in S, t \not\in S$. So for any $s \in S$ and for any $t \equiv s$ we have $t \in S$, i.e.~$S$ is a union of some equivalence classes of $\equiv$. Call this set of classes $X$ (so in the earlier notation, $S = \cup X$). By \eqref{eq:footz}, in $G / \!\equiv$ (which is $\gcac$) we have $\delta(X) = \{e, f\}$, so $\{e, f\}$ is a cut pair of $\gcac$ as needed.
\end{proof}
Now we recall the earlier convention (from \prettyref{sec:randcirc}) of using $\phi$ to denote a (random) $b$-bit binary circulation, wherein for each $i$ the sets $\{e \mid \phi_i(e)=1\}$ are independent binary circulations selected uniformly at random.
Recall also \prettyref{cor:seppair} which (together with the assumption that $G$ is 2-edge-connected) says that $\phi(e)=\phi(f)$ always holds when $\{e, f\}$ is a cut pair, and holds with probability $1/2^b$ otherwise. Define an \emph{illusory cut pair} to be a pair of edges $\{e, f\}$ that has $\phi(e)=\phi(f)$ but is not a cut pair.
Our strategy for the parallel algorithm will be to define another relation $\equiv'$ so that $\equiv$ and $\equiv'$ will agree when there are no illusory cut pairs;
we will then use \prettyref{prop:cpp} to verify that there are no illusory cut pairs.

\subsubsection{Cactuslike Graphs}
The relation $\equiv'$ must provide an alternate way of constructing $\gcac$, when there are no illusory cut pairs. To this end we examine the properties of $\gcac$ in more detail.
We fix some terminology: a \emph{closed walk} has distinct edges but may repeat vertices; a \emph{simple cycle} is any closed walk without repeated vertices. Thus a simple cycle of length 1 is a loop, and a simple cycle of length 2 is a parallel pair of non-loop edges.

An \emph{ear decomposition} of $G$ is a sequence of graphs $G_0 \subset G_1 \subset G_2 \subset \dotsb \subset G_k = G$ such that $G_0$ is just a vertex and each $G_i$ is obtained from $G_{i-1}$ by attaching a simple cycle (a \emph{closed ear}) or path with both endpoints in $G_{i-1}$ (an \emph{open ear}). It is well-known that a graph is 2-edge-connected if and only if it admits an ear decomposition. The path or cycle added to $G_{i-1}$ to get $G_i$ is called the $i$th \emph{ear} of $G$, and it is denoted $E_i$.

We omit the straightforward proof of \prettyref{prop:2c2c}.
\begin{prop}\label{prop:2c2c}
Every pair of nodes in $\gcac$ has edge-connectivity equal to 2.
\end{prop}
Call a graph \emph{cactuslike} if every pair of nodes has edge-connectivity equal to 2.
\begin{prop}\label{prop:2cactus}
The following are equivalent for any graph: (a) it is cactuslike; (b) in some ear decomposition, all its ears are closed; (c) in every ear decomposition, all its ears are closed; (d) every edge lies in exactly one simple cycle.
\end{prop}
\begin{proof}
Trivially, (c) implies (b). It is easy to see that (b) implies (a) by induction on the ears.

We now prove the contrapositive of (a) $\Rightarrow$ (d). If (d) is false there are two simple cycles $C_1, C_2$ both containing an edge $e$; since the $C_i$ are different there is another edge $f$ with $f \in C_1, f \not\in C_2$ (WOLOG). Let $P$ be the inclusion-minimal subpath of $C_1$ containing $f$ with both its endpoints in $C_2$. Then we get three edge-disjoint paths connecting the endpoints of $P$: $P$ itself plus two in $C_2$. By the definition of cactuslike, (a) is false, and we are done.

We also prove the contrapositive of (d) $\Rightarrow$ (c). If (c) is false there is an ear decomposition with an open ear $E_i$ with endpoints $u, v$. Since $G_{i-1}$ is 2-edge-connected, there are two different simple $u$-$v$ paths in $G_{i-1}$. Combining these paths in turn with $E_i$ gives two simple cycles having at least one common edge, so (d) is false.
\end{proof}
From \prettyref{prop:2cactus}(c) we obtain the following corollary by induction on the ears.
\begin{cor}\label{cor:ccp}
In a cactuslike graph, for any ear decomposition, the cut classes are the same as the nonsingleton ears.
\end{cor}

We know from \prettyref{prop:2c2c}, along with the definitions of $\gcac$ and cactuslike, that $\gcac$ is a cactuslike contraction of $G$, and by \prettyref{prop:cpp} the cut classes of $\gcac$ are the same as the cut classes of $G$. The following converse will be very useful.
\begin{lmma}\label{lmma:2way}
Let $R$ be an equivalence relation for which $G/R$ is cactuslike, and such that the cut classes of $G/R$ and the cut classes of $G$ are the same. Then $R$ is the same as $\equiv$.
\end{lmma}
\begin{proof}
First, we establish that $\equiv$ refines $R$. Suppose otherwise, that there are vertices $u, v$ related by $\equiv$ but not by $R$. Then the vertices corresponding to the equivalence classes of $u$ and $v$ in $G/R$ are different. Since $G/R$ is cactuslike, there is a cut pair separating those vertices. But then \eqref{eq:footz} shows this cut pair also separates $u$ from $v$ in $G$, contradicting the fact that $u \equiv v$.

Now, we establish that $R$ refines $\equiv$. Suppose otherwise, that there are vertices $u, v$ related by $R$ but not by $\equiv$. Since $u \nequiv v$ there is a cut pair $\{e, f\}$ in $G$ separating $u$ from $v$. Thus $G \bs \{e, f\}$ has two connected components, one containing $u$ and one containing $v$, and since $u R v$, we see $G/R \bs \{e, f\}$ is connected. So $G$ and $G/R$ have different cut pairs, hence different cut classes, which is the needed contradiction.
\end{proof}

\subsubsection{Pinching Ears}
In this subsection we develop an algorithmic ear-based tool to construct $\equiv$. First we need the following.
\begin{lmma}\label{lmma:ecc}
Every cut class of $G$ lies within a single ear of any ear decomposition of $G$.
\end{lmma}
\begin{proof}
Suppose otherwise, that there is a cut pair $\{e, f\}$ with $e \in E_i$ and $f \in G_{i-1}$. Since $G_{i-1}$ is 2-edge-connected, $G_{i-1} \bs f$ is connected. But $G_i \bs \{e, f\}$ is obtained by attaching 2 paths to $G_{i-1}$, and so is connected. By induction on the remaining ears we see that each $G_j \bs \{e, f\}$ for $j \geq i$ is connected; in particular for $j = k$ this means $G \bs \{e, f\}$ is connected, a contradiction.
\end{proof}
The cut classes of $\gcac$ and $G$ agree (\prettyref{prop:cpp}), and by \prettyref{cor:ccp} the cut classes of $\gcac$ are the same as its non-loop simple cycles. We reiterate for future reference:
\begin{equation}\label{eq:gg}
\textrm{The cut classes of $G$ are the same as the non-loop simple cycles in $\gcac$.}\tag{\ensuremath{\diamondsuit}}
\end{equation}

We now define a \emph{pinching} operation whose effect is also to turn a cut class into a cycle.
Let a given ear $E_i$ have vertices and edges $v_0, e_1, v_1, e_2, v_2, \dotsc, e_z, v_z$ in that order, where $v_0 = v_z$ iff the ear is closed. For a given subset $U = \{e_{j(1)}, e_{j(2)}, \dotsc, e_{j(t)}\}$ of the ear's edges indexed such that $j(1) < j(2) < \dotsb < j(t)$, let $R_U$ denote the equivalence relation on the ear's vertices consisting exactly of the pairs
$$R_U := \bigl\{\{v_{j(1)}, v_{j(2)-1}\}, \{v_{j(2)}, v_{j(3)-1}\}, \dotsc, \{v_{j(t-1)}, v_{j(t)-1}\}, \{v_{j(t)},  v_{j(1)-1}\}\bigr\}.$$ Thus in the ``pinched" ear $E_i / R_U$, the set $U$ forms a simple cycle. For example if $R_U$ consists of just one edge, then that edge becomes a loop in $E_i / R_U$.

\begin{defn}
Denote the cut classes contained in ear $i$ as $U[1], U[2], \dotsc, U[s]$, then define $\equiv_i$ to be the equivalence relation defined by the transitive closure $(R_{U[1]} \cup R_{U[2]} \cup \dotsb \cup R_{U[s]})^*$.
\end{defn}
In other words, $\equiv_i$ simultaneously pinches all cut classes appearing in ear $E_i$.

\begin{lmma}\label{lmma:refinally}
For each $i$, $\equiv_i$ refines $\equiv$.
\end{lmma}
\begin{proof}
Using \prettyref{lmma:ecc} and the preceding definitions, it is necessary and sufficient to show that for each cut class $U$, every pair of vertices related by $R_U$ is also related by $\equiv$.

It is not hard to see that we may use $U$ to decompose $G$ into $|U|$ 2-edge-connected graphs linked in a cycle by edges of $U$. We illustrate this on the left side of Figure~\ref{fig:refinally}; compare with \prettyref{fig:cc}. Moreover, it is clear that the order of appearance of $U$ in $E_i$ is the same as the cyclic order of $U$ in this decomposition. Therefore, adopting the notation in the definition of $R_U$, we label the 2-edge-connected graphs $G_1, G_2, \dotsc, G_{|U|}$ such that edge $e_{j(x)}$ joins $G_{x-1}$ to $G_x$ (or if $x=1$, $G_{|U|}$ to $G_1$). The notation is illustrated on the right side of Figure \ref{fig:refinally}.

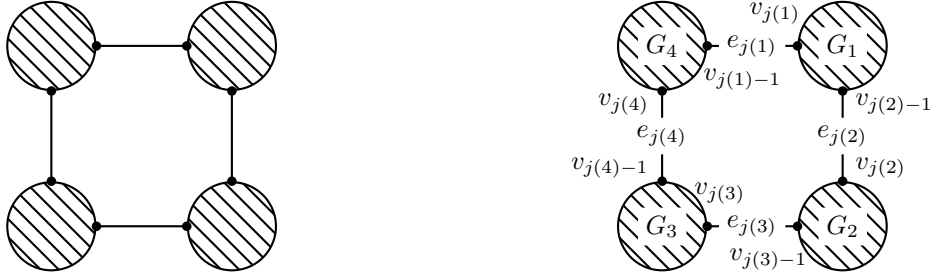
\begin{figure}[t]
\begin{center} \leavevmode
\begin{pspicture}(-2,-2)(4,2)
  \psset{unit=0.4}
  \psellipse[fillstyle=vlines](3,3)(1.5,1.5)
  \psellipse[fillstyle=vlines](-3,3)(1.5,1.5)
  \psellipse[fillstyle=vlines](-3,-3)(1.5,1.5)
  \psellipse[fillstyle=vlines](3,-3)(1.5,1.5)
  \psline[arrows=*-*](-1.5,3)(1.5,3)
  \psline[arrows=*-*](-1.5,-3)(1.5,-3)
  \psline[arrows=*-*](3,-1.5)(3,1.5)
  \psline[arrows=*-*](-3,-1.5)(-3,1.5)
\end{pspicture}
\begin{pspicture}(-4,-2)(2,2)
  \psset{unit=0.4}
  \psellipse[fillstyle=vlines](3,3)(1.5,1.5)
  \psellipse[fillstyle=vlines](-3,3)(1.5,1.5)
  \psellipse[fillstyle=vlines](-3,-3)(1.5,1.5)
  \psellipse[fillstyle=vlines](3,-3)(1.5,1.5)
  \psline[arrows=*-*](-1.5,3)(1.5,3)
  \psline[arrows=*-*](-1.5,-3)(1.5,-3)
  \psline[arrows=*-*](3,-1.5)(3,1.5)
  \psline[arrows=*-*](-3,-1.5)(-3,1.5)
  \rput*(-3,3){$G_4$}
  \rput*(3,3){$G_1$}
  \rput*(3,-3){$G_2$}
  \rput*(-3,-3){$G_3$}
  \uput[-30](3,1.5){$v_{j(2)-1}$}
  \uput[30](3,-1.5){$v_{j(2)}$}
  \rput*(3,0){$e_{j(2)}$}
  \uput[-210](-3,-1.5){$v_{j(4)-1}$}
  \uput[210](-3,1.5){$v_{j(4)}$}
  \rput*(-3,0){$e_{j(4)}$}
  \rput*(0,-3){$e_{j(3)}$}
  {\psset{labelsep=0.75}
  \uput[75](-1.5,-3){$v_{j(3)}$}
  \uput[-115](1.5,-3){$v_{j(3)-1}$}}
  \rput*(0,3){$e_{j(1)}$}
  {\psset{labelsep=0.75}
  \uput[115](1.5,3){$v_{j(1)}$}
  \uput[-60](-1.5,3){$v_{j(1)-1}$}}
\end{pspicture}
\end{center}
\caption{Left: the structure of a 2-edge-connected graph with respect to a cut class (shown with a cut class of size 4); each hatched region indicates a 2-edge-connected subgraph (possibly a single vertex). Right: the notation used in the proof of \prettyref{lmma:refinally}.} \label{fig:refinally}
\end{figure}

With this setup, we now show that every pair of edges related by $R_U$ is also related by $\equiv$. First, $v_{j(x)} \equiv v_{j(x+1)-1}$ for $1 \le x < |U|$ since these two vertices are 2-edge-connected in $G_x$ and also are linked by a third disjoint path going the long way around the cycle (using all of $U$). Establishing that $v_{j(t)} \equiv v_{j(1)-1}$ is similar.
\end{proof}

\begin{prop} The equivalence relation $(\cup_i \equiv_i)^*$ is the same as $\equiv$.\label{prop:tar}
\end{prop}
\begin{proof}
This is where we make use of \prettyref{lmma:2way}. The main part of the proof is to show that the simple cycles in $G/(\cup_i \equiv_i)^*$ are the same as the simple cycles in $G / \mathord{\equiv}$ (which by \eqref{eq:gg} are the same as $G$'s cut classes). Supposing we can prove this, it follows by two applications of \prettyref{prop:2cactus}(d) that $G/(\cup_i \equiv_i)^*$ is cactuslike, then \prettyref{cor:ccp} shows that $G/(\cup_i \equiv_i)^*$ has the same cut classes as $G$, then applying \prettyref{lmma:2way} we are done.

Using \prettyref{lmma:ecc} any cut class $U$ of $G$ becomes a simple cycle in $G/R_U$ and hence a closed walk in $G/(\cup_i \equiv_i)^*$ (because that graph is a contraction of $G/R_U$). Moreover, we know by \eqref{eq:gg} that every cut class of $G$ is a simple cycle of $G/\mathord{\equiv}$, and since $(\cup_i \equiv_i)^*$ refines $\equiv$, it follows that every cut class of $G$ is becomes a simple cycle in $G/(\cup_i \equiv_i)^*$. Reiterating, every cut class of $G$ is a simple cycle of $G/(\cup_i \equiv_i)^*$.

Could there be any other simple cycles in $G/(\cup_i \equiv_i)^*$ --- one which is not just a cut class of $G$? We will show the answer is no, which will complete the proof. Suppose that $C$ is any simple cycle in $G/(\cup_i \equiv_i)^*$. Since $(\cup_i \equiv_i)^*$ refines $\equiv$, we may view $G/\mathord{\equiv}$ as a contraction of $G/(\cup_i \equiv_i)^*$, thus the image of $C$ in $G/\mathord{\equiv}$ is a closed walk. But any closed walk is an edge-disjoint union of simple cycles (this is just a decomposition of an Eulerian graph), and the simple cycles in $G/\mathord{\equiv}$ are the cut classes of $G$; so $C$ is a union of cut classes of $G$. But the previous paragraph establishes each cut class of $G$ becomes a simple cycle in $G/(\cup_i \equiv_i)^*$; and in $G/(\cup_i \equiv_i)^*$ it is impossible for $C$ to be simultaneously a single simple cycle and a union of more than one simple cycle. Thus $C$ is just a single cut class of $G$, as needed.
\end{proof}

\subsubsection{Detecting Errors}
In the algorithm we are designing, we don't know the cut pairs; rather, we have computed $\phi$ and know that with high probability, $\phi$ labels edges by their cut class. We compute the following instead.
\begin{defn}
For the $i$th ear $E_i$, enumerate $\{\phi(e) \mid e \in E_i\}$ as $\{x_1, x_2, \dotsc, x_s\}$. Let $W[k]$ denote the set $\{e \in E_i \mid \phi(e) = x_k\}$ and define $\equiv'_i$ to be the equivalence relation defined by the transitive closure $(R_{W[1]} \cup R_{W[2]} \cup \dotsb \cup R_{W[s]})^*$.
\end{defn}
In other words, $\equiv_i$ simultaneously pinches all sets of common $\phi$-value in ear $E_i$.

Note first that if there are no illusory cut pairs, then the sets of common $\phi$-value are the same as the cut classes, and so $\equiv'_i$ is the same as $\equiv_i$. Define the equivalence relation $\equiv'$ to be equal to $(\cup_i \equiv'_i )^*.$

\begin{thm}\label{thm:lvpcv}
There is a Las Vegas parallel algorithm to compute all cut pairs in $O(\log V + T(E))$ time, $O(E+S(E))$ space, and
$O(E+W(E))$ work, in expectation.
\end{thm}
\begin{proof}
Our algorithm computes $H := G / \mathord{\equiv'}$ and tries to verify all cut pairs. To compute the relations $R_{W[i]}$ on each ear, we sort the edges on that ear lexicographically according to the pair $(\phi(e), pos(e))$ where $pos(e)$ is the position along the ear. Then to compute the transitive closure $\equiv'$ of their union, we build an auxilliary graph on vertex set $V$ and draw an edge for each pair of vertices that is related by some $R_{W[i]}$ on some ear; then the equivalence classes of $\equiv'$ are the connected components of this auxilliary graph. This can be done using the connected components routine of \cite{HZ96}. From this, computing the contracted multigraph $H$ takes constant time and linear work.

Now we check if $H$ is cactuslike, by computing an ear decomposition and seeing if all ears are closed. First, if $H$ is not cactuslike, by \prettyref{prop:tar}, the verifier can reject since there is an illusory cut pair (since $\equiv \neq \equiv'$). Second, if $H$ is cactuslike, then its cut classes are the same as the ears. Using this fact, and sorting all edges by their $\phi$ value, the verifier accepts iff every pair $\{e, f\}$ with $\phi(e)=\phi(f)$ is a cut pair of $H$. By \prettyref{prop:cpp}(a), the verifier will reject when there is an illusory cut pair, and accept otherwise.
\end{proof}

\section{Future Work} At the most basic level, it would be interesting to push further and find efficient algorithms for higher types of connectivity, such as finding all 3-edge-cuts in $O(E)$ sequential time or $O(\Diam)$ distributed time. The state of the art for this problem in the sequential model is $O(V^2)$ time \cite{galil-ital,KR91}. It would also be interesting to reduce the complexity of our parallel cut pairs algorithm to linear work and logarithmic time; it seems plausible that another approach would avoid radix sort.

It is possible to deterministically compute the cut edges in the distributed model using $O(\Diam)$ time and $O(E)$ messages, as was shown in the thesis \cite{dp-thesis}. (The approach is based on the observation that $\{v, p(v)\}$ is a cut edge if and only if $low(v) \geq v$ and $high(v) < v + desc(v)$.) However, we do not know of any deterministic analogues of our distributed cut pair or cut vertex algorithms.

It would be interesting to know if
our distributed cut vertex algorithm could be synthesized with the
cut vertex algorithm of \cite{Thur97} to yield further
improvement. Alternatively, a lower bound showing that no $O(\Diam)$-time algorithm is possible for finding cut vertices would be very interesting.

\subsection*{Acknowledgments}We would like to thank Graeme Kemkes, Jochen K\"onemann, and the referees from ICALP and ACM Trans.~Alg.~for valuable feedback.

\bibliography{../huge}
\bibliographystyle{alpha}

\end{document}